\documentclass[a4paper,11pt]{article}
\usepackage{amsmath}
\usepackage{amsfonts}
\usepackage{amssymb}
\usepackage{array}
\usepackage{bm}
\usepackage{censor}
\usepackage{colortbl}
\usepackage[utf8x]{inputenc}
\usepackage{ae}
\usepackage{enumitem}
\usepackage[T1]{fontenc}
\usepackage[english]{babel}
\usepackage{graphicx}
\newcommand{\indep}{\rotatebox[origin=c]{90}{$\models$}}
\usepackage{makecell}
\usepackage{placeins} % For \FloatBarrier
\usepackage{siunitx}
\usepackage{tabularx}
\usepackage{tabularray}
\usepackage{tabulary}
\usepackage{setspace}
\usepackage{soul} % for highlighting text, using \sethlcolor{lightgray} abd \hl{This is shaded text.}
\usepackage{tablefootnote}
\usepackage[flushmargin]{footmisc}
\usepackage[comma]{natbib}
\usepackage{float}
\usepackage{caption}
\usepackage{tikz}
	\usetikzlibrary{trees}
	\usetikzlibrary{shapes.geometric, arrows.meta, positioning, calc}
	\usetikzlibrary{decorations.pathreplacing}
\usepackage[a4paper,left=2.5cm,right=2.5cm,top=3cm,bottom=3cm]{geometry}
\usepackage[colorinlistoftodos]{todonotes}
\usepackage{pdfpages}
\usepackage{longtable}
\usepackage{adjustbox}
\usepackage{booktabs}
\usepackage{multirow}
\usepackage{lscape}
\usepackage[colorlinks=true, allcolors=blue]{hyperref}

\usepackage[flushleft]{threeparttable}
\usepackage{inputenc}
\usepackage{xcolor}



\begin{document} \doublespacing \pagestyle{plain}
	
	\def\ci{\perp\!\!\!\perp}
	\begin{center}
		
		{\LARGE How (Not) to Incentivize Sustainable Mobility? \\ Lessons from a Swiss Mobility Competition}

		{\large \vspace{0.8cm}}
		
		{\large Silvio Sticher, Hannes Wallimann and Noah Balthasar }\medskip

		{\small {University of Applied Sciences and Arts Lucerne, Institute of Tourism and Mobility} \bigskip }
		
		{\large \vspace{0.8cm}}
		
		{\large Version August 2024}\medskip

	\end{center}
	
	\smallskip

	\noindent \textbf{Abstract:} {
		We investigate the impact of a gamified experiment designed to promote sustainable mobility among students and staff members of a Swiss higher-education institution. Despite transportation being a major contributor to domestic CO\textsubscript{2} emissions, achieving  behavioral change remains challenging. In our two-month mobility competition, structured as a randomized controlled trial with a 3$\times$3 factorial design, neither monetary incentives nor norm-based nudging significantly influences mobility behavior. Our (null) results suggest that there is no "gamified quick fix" for making mobility substantially more sustainable. Also, we provide some lessons learned on how \emph{not} to incentivize sustainable mobility by addressing potential shortcomings of our mobility competition.}
%Transportation is responsible for around 40\% of CO\textsubscript{2} emissions in Switzerland. To reduce CO\textsubscript{2} emissions, it is therefore important to investigate sustainable changes in mobility behavior. The paper presents a experiment using the gamification approach, which aims to promote sustainable mobility through a challenge. Participants voluntarily record their mobility patterns using an app. The aim is to positively influence mobility behavior through communication of monetary incentives and social norms. The evaluation shows that the experimental groups do not significantly differ from each other in terms of the points achieved in the challenge, based on the means of transportation used. Overall, however, the points and corresponding emissions increase in all groups at weekends and on vacations. The collected data provide valuable insights into the effectiveness of various communication strategies. The findings contribute to the understanding of the gamification approach in the transportation sector, including its potential and limitations to promote environmentally sustainable mobility practices.
	{\small \smallskip }
	{\small \smallskip }
	{\small \smallskip }
	
	
	{\small \smallskip }
	{\small \smallskip }
	{\small \smallskip }
	
	{\small \noindent \textbf{Acknowledgments:} We are grateful to the Albert Köchlin Stiftung, Luzernmobil, Caritas Luzern, Die Zentralbahn, CKW, PostAuto, and Verkehrsbetriebe Luzern for their financial support. We have benefited from various inputs, including those from Ruben Antenen, Chiara Köchli, Andreas Merz, Timo Ohnmacht, Sarah Troxler, and Widar von Arx.}
	
	\bigskip
	\bigskip
	\bigskip
	\bigskip
	
	{\small {\scriptsize 
\begin{spacing}{1.5}\noindent  
\textbf{Addresses for correspondence:} Silvio Sticher, University of Applied Sciences and Arts Lucerne, Rösslimatte 48, 6002 Lucerne, \href{mailto:silvio.sticher@hslu.ch}{silvio.sticher@hslu.ch}; Hannes Wallimann, \href{mailto:hannes.wallimann@hslu.ch}{hannes.wallimann@hslu.ch}; Noah Balthasar, \href{mailto:noah.balthasar@hslu.ch}{noah.balthasar@hslu.ch}.
\end{spacing}
			
		}\thispagestyle{empty}\pagebreak  }

	{\small \renewcommand{\thefootnote}{\arabic{footnote}} %
		\setcounter{footnote}{0}  \pagebreak \setcounter{footnote}{0} \pagebreak %
		\setcounter{page}{1} }
	
\section{Introduction}\label{introduction}

As part of the Paris Climate Agreement, the international community has set the goal of limiting global warming to a maximum of 2 degrees Celsius by 2100 and reducing greenhouse gas emissions by 40\% by 2050 \citep{IPCC2023}. In Switzerland, transportation accounts for approximately 33\% of domestic CO\textsubscript{2} emissions \citep{bafu_co2_statistics}.\footnote{Strictly speaking, this figure refers to greenhouse gases (CO\textsubscript{2} equivalents). For simplicity, we refer to these as CO\textsubscript{2} emissions throughout the article.} Compared with other sectors, in transportation, absolute CO\textsubscript{2} numbers remain stubbornly high and, as \cite{creutzig2014transport} note, will continue to rise as long as behavioral changes, technological advancements, and modifications in the built environment fail to decouple emissions from economic activity. Due to the slow technological transition in transportation, there is a need to accelerate a mobility transition, which entails changing individual behavior. 

Mobility behavior can be influenced by a range of measures, including infrastructure modifications, economic conditions, monetary incentives, and social norms. While "push measures" that urge consumers to use environmentally friendly means of transportation are considered relatively effective \citep{hekler2022push}, they often face significant challenges in implementation due to political obstacles. Hence, and with the increasing possibilities provided by digital tools, highly accepted "pull measures", including such that reduce the informational burden for climate-friendly travel (such as Mobility-as-a-Service platforms) and rewards-based incentives, have gained ground \citep{alatacs2021role}. Information and communication technologies thereby facilitate the inclusion of playful elements in an approach called "gamification." In several cases, as in the one we present in this article, passive tracking technology is used to record journeys; patterns, including the most probable means of transportation, are derived; and these patterns are used to inform, persuade, and reward desired behaviors in an attempt to be both effective and create a positive experience simultaneously. 

In our case, we engaged 264 students and staff members of the Lucerne University of Applied Sciences in Switzerland (HSLU) to participate in a two-month-long 'mobility competition' in March and April 2024, using the Motiontag tracking app. Of the 264 registered, 195 both participated and had their data retained in compliance with data subject rights. \emph{Information} (about desired and actual behavior) was provided by allocating points to the preferred means of transportation, including cycling, walking, and public transportation, and regularly informing participants about their preliminary totals. \emph{Persuasion} was implemented by framing the information based on either descriptive or injunctive social norms. \emph{Rewarding} took place through monetary incentives in the form of competition prizes, leveraging insights from prospect theory \citep{kahneman1979prospect}. To account for likely self-selection, we employed an experimental approach with random assignment of multiple treatments in terms of incentives and communication, resulting in a 3$\times$3 factorial design (including control groups regarding both incentives and communication). 

Given the total prize sum of CHF\,3,150 (roughly US\$\,3,600), our main result may come as a surprise. Neither persuasion nor reward had any measurable impact on mobility behavior when measured as the individual sum of points collected. Two different but potentially complementary explanations come to mind: Either mobility, including the choice of means of transportation, is rather inelastic in a wealthy country such as Switzerland, at least in the short term, or this is rather a lesson in how \emph{not} to set up a mobility competition. In the article at hand, we explore both possibilities.

By revealing limitations of gamified and reward-based interventions in a mobility competition, our study offers insights that caution against over-reliance on such strategies. Our findings provide a counterbalance to the optimism surrounding digital tools in behavioral change, highlighting the need for more nuanced, context-specific approaches to reducing mobility-related CO\textsubscript{2} emissions. Consequently, our work may serve as a reference point for policymakers and practitioners aiming to implement effective and evidence-based solutions for sustainable mobility.

We structure the remainder of our paper as follows. In Section \ref{literature}, we provide a literature overview of the current state of research on gamification in the field of sustainable mobility, with a focus on incentives and communication. In Section \ref{methods}, we present our experimental design, which includes both the setup of the mobility competition and our methodology to assess causal effects. We then show our descriptive and inferential results in Section \ref{results}. We discuss our results (or the lack thereof) in Section \ref{discussion} and conclude in Section \ref{conclusion}.

\section{Literature}\label{literature}

In recent years, mobility\footnote{When speaking of "mobility" in lieu of "transportation", we emphasize geographic mobility resulting from individual behavior. "Mobility" relates to the overall experience of movement in "possibility spaces", whereas "transportation" focuses on the measurable realization of this movement \citetext{\citealp[p. 117]{canzler1998moglichkeitsraume}; \citealp[p. 42]{ohnmacht2008mobilitatsbiografie}}.} studies have shown growing interest in using "gamified" approaches to nudge behaviors toward being less environmentally harmful. While game elements are employed to tackle an array of goals, including the promotion of road safety and socially compatible mobility, most measures aim to shift user behavior towards more CO\textsubscript{2}-efficient practices. Studies have explored how apps and games can be used to track data, deliver messages, and promote awareness of transportation-related carbon emissions and alternative modes of transportation, influencing mobility behavior without resorting to hard measures such as imposing restrictions, bans, or spatial modifications \citetext{see, e.g., \citealp{cellina2019large}; \citealp{ferron2019play}; \citealp{anagnostopoulou2020mobility}}.

It is repeatedly shown that gamified approaches---primarily favoring public transportation, micromobility, cycling, walking, and car sharing---can have a positive impact on people's travel behavior from an environmental perspective \citep[for an overview, see][]{wang2022initiatives}.\footnote{Note that in this literature review, we focus specifically on gamification in the context of mobility. For a meta-analysis of gamification across various domains, see, for example, \cite{hamari2014does}.} \cite{butnaru2020european} emphasizes the potential of "green gamification" as part of the European Green Deal, particularly in the transportation sector, to reduce the use of private vehicles through engaging and interactive games. \cite{schillaci2022gamification} investigate the effectiveness of a mobile app to promote sustainable commuting through gamification and point toward significant CO\textsubscript{2} reductions. \cite{bui2015effects} examine the impact of a gamified information system in a car-sharing service and show how gamification can change driver behavior to reduce CO\textsubscript{2} emissions. The study by \cite{minnich2023gamification} examines the effect of gamified campaigns on individual mobility behavior in Hamburg, Germany, employing a quasi-experimental setting. He finds that participants affected by these campaigns increased their cycling distances. Similarly, \citet{weber2018convergence} investigate how changes in a "bicycle promotion game" in the UK, Australia, and the US affected the frequency of bicycle trips. In another example, an analysis in Bologna, Italy, shows that 47\% of participating motorists exhibited sustained engagement and behavioral change due to game incentives \citep{bowden2019data}. 

Ideally, gamification enables new experiences that lead to a reflection on habitual behavior patterns. To catalyze awareness into action, both monetary incentives and persuasive communication are commonly used, as we do in our study.

%\cite{papaioannou2018iot} present a gamified app to promote energy saving in the workplace, emphasizing user-centered design and positive user engagement. 

%The use of gamification in the transportation sector aims to learn, enforce, and change mobility-related behaviors. Behavioral change is achieved voluntarily and not as a result of spatial modifications. 

\paragraph{Incentives}
It is common sense and well-documented that prices affect mobility behavior.\footnote{A meta-analysis in public transportation shows that the short-run price elasticity (with vehicle kilometers treated as exogenous) is $\epsilon = -0.38$ \citep{holmgren2007meta}. Similarly, regarding fuel prices, the average short-run price elasticity in motorized private transportation is estimated to be $\epsilon = -0.34$ \citep{brons2008meta}. In Switzerland, a recent experimental study estimates the short-term elasticity of external costs in reaction to price increases to be $\epsilon = -0.24$ \citep{hintermann2024pigovian}.} However, the lack of political and societal acceptance may jeopardize efforts to shift mobility towards more sustainable forms through price interventions. Consequently, there is a strong preference for strategies that achieve similar outcomes through voluntary means, specifically by rewarding desirable behavior.

Among the measures explored are flat incentives to encourage cycling. \cite{maca2020incentivizing} demonstrate in their randomized field experiment that providing a small monetary reward for each kilometer cycled to work or school significantly increased commuter cycling in Czech cities. This result is mirrored by \cite{ciccone2021using}, who conducted a similar experiment in Norwegian cities. In addition to flat incentives, they introduced a lottery version of the incentive scheme (similar to our treatment \emph{Constant incentives}, see Section \ref{methods}), which yielded slightly less promising but still significantly positive results. Regarding public transportation, \cite{gravert2021nudges} show in a large-scale study in Sweden that economic incentives in the form of free trial periods can have lasting effects through habit formation.

\paragraph{Communication}

In the same study, however, \cite{gravert2021nudges} also show that the mere provision of informational nudges is less likely to be effective. Such "suasoric" measures, which rely on invoking social expectations (descriptive norms) or moral arguments (injunctive norms) to influence behavior, are characterized by their low-threshold nature and cost efficiency. They are often seen as a complement to economic instruments. 

%The literature on mobility behavior offers numerous articles investigating the success of applying these psychological findings to the transportation sector through "persuasive technology," such as apps \citetext{\citealp[see, e.g.,][]{fogg2002persuasive}; %"as though the computers were social entities
%\citealp{froehlich2009ubigreen}; %13 participants, feedback, participants did start new behaviours
%\citealp{jylha2013matkahupi}; %persuasive: of 149 challenges, 105 were accepted, 95 completed
%\citealp{jariyasunant2015quantified}}. % automated diary system --> significance of peer feedback, driving distances went back by 33%

Some studies have found promising results when applying psychological insights to influence mobility behavior through "persuasive technology," such as mobile applications. For example, the formative work of \cite{froehlich2009ubigreen} demonstrates that providing feedback on transportation choices can lead to increased engagement and the adoption of new behaviors (though their study was limited to just 13 participants). Similarly, \cite{jylha2013matkahupi} exemplify the benefit of persuasion by challenges, and \cite{fogg2002persuasive} confirms that individuals tend to respond to computer systems in ways that reflect their interaction with social entities in terms of motivation and influence. Further support for the potential of such instruments comes from \cite{jariyasunant2015quantified}, who show that feedback and peer influence contributed to a significant and substantial reduction in driving distances in a three-week experiment in San Francisco.

The broader evidence regarding the effectiveness of communication-based strategies for changing mobility behavior, however, remains mixed at best.
\cite{sunio2017promote}, for instance, state that methodologically robust analyses are lacking and that no conclusive assessments should be made. The study by \cite{cellina2019large},
%\textcolor{red}{Kleines Detail: \cite{sunio2017promote} werden in der Aufzählung nicht genannt, \citealp{cellina2019large} schon. Hat das einen Grund? Wenn nein, würde ich es einheitlich machen.}
which, like our study, discusses a randomized controlled trial in Switzerland, found no significant overall effects from adding persuasive elements (such as the possibility of chasing individual goals) to a mobility-tracking app. Summing up five high-powered field experiments which failed to shift commuter behavior, \cite{kristal2020we} suggest that suasoric communication---with nudges "ranging from simple framing interventions to more resource-intensive [...] interventions [...] such as signing up for an in-person meeting"---is largely ineffective in this area.\footnote{Looking at our own study, it is worth mentioning that \cite{kristal2020we} also stress the importance of publishing null results.} 
%Cellina: Intrusive tracking, dropout and a potentially unrepresentative sample lead to uncertainties in prediction.

\section{Methods and Experimental Setting}\label{methods}

The aim of our study is to identify the causal effects of treatments, including monetary incentives, persuasive communication, and---hoping to find complementarities---their combinations. After outlaying our identification strategy in Subsection \ref{identification} and defining our outcome variable in Subsection \ref{scores}, we present the experimental setting (mobility competition) in greater detail in Subsection \ref{setting}, covering recruitment, group allocation based on the factorial design, tracking and trip construction, and a detailed specification of the treatments.

\subsection{Identification Strategy}\label{identification}

Based on the potential outcome framework of \citet{rubin1974estimating}, we denote the potential outcomes $Y(1)$ and $Y(0)$ as the outcomes $Y$ ("scores" describing mobility behavior, see Subsection \ref{scores}) that are realized or would be realized if the binary treatment $D$ (regarding incentives, communication, and interactions) takes on the value $D=1$ or $D=0$, respectively. Because treatment is exogenously assigned in this randomized controlled trial, no characteristics systematically and erroneously link $D$ to $Y$; thus, $Y(1)-Y(0)$ represents the causal effect of $D$ on $Y$. Since we only observe either $Y(1)$ or $Y(0)$ at an individual level, we measure causal effects on an aggregate level, specifically as average treatment effects (ATEs), formally defined as $\text{ATE}=\mathbb{E}[Y(1)-Y(0)]=\mathbb{E}[Y|D=1]-\mathbb{E}[Y|D=0]$.

We use linear regression to calculate ATEs by fitting a straight line: $Y=\beta_0+\beta_1D+\beta_2X_2+\dots+\beta_kX_k+\varepsilon$. Here, $\beta_0$ represents the mean outcome of the control group ($\mathbb{E}[Y|D=0]$), $\beta_1$ estimates the ATE, and $\varepsilon$ is the error term. The variables $X_2$ to $X_k$, with coefficients $\beta_2$ through $\beta_k$, control for additional treatments or background information (specifically whether the individual is a student or staff member) to improve estimate precision.

%... and $\varepsilon$ to the error term (according to the first-moment condition $\mathbb{E}[\varepsilon]=0$)
%To estimate the ATE in our experiment, we use the statistical software \textsf{R}. 

The identification of ATEs using counterfactuals $Y(0|D=1)$ and $Y(1|D=0)$ relies on two key assumptions \citep[see, e.g., ][]{huntington-klein2021effect}. 
\begin{enumerate}
	\item The \emph{Independence Assumption}, requiring random assignment of $D$, is satisfied by design, as detailed in Subsection \ref{setting}. This ensures that potential outcomes are independent of treatment status. Formally, $\{Y(1),Y(0)\}\indep D$.
	\item The \emph{Stable Unit Treatment Value Assumption} (SUTVA) stipulates that individual potential outcomes are unaffected by treatments of other participants. Differently put, we assume no treatment spillover effects. Given that our participants are spread across six departments, most with multiple locations, SUTVA also seems plausible.
\end{enumerate}

\subsection{Definition of the Outcome Variable}\label{scores}

We placed "scores" at the center of both our incentive structure and participant communication. Points were awarded based on the main means of transportation (as discussed in Subsection \ref{setting}) for each recorded trip. Trips using active means of transportation earned two points. Public transportation trips were awarded one point, while trips using motorized individual transportation received zero (but not negative) points.

In Table \ref{tab:scores}, we display the main modes that were automatically detected during the mobility competition, along with the points allocated.  
	
\begin{table}[H]
	\centering
	\footnotesize
	\singlespacing	
	\renewcommand{\arraystretch}{1.1}
	\begin{threeparttable}	
		\caption{Allocation of points based on means of transportation (main means)}\label{tab:scores}
	\begin{tabular}{>{\centering\arraybackslash}p{4.5cm}>{\centering\arraybackslash}p{3.5cm}>{\centering\arraybackslash}p{3.5cm}} 
%			\begin{tabular}{ccc} 
			\toprule 
			 Motorized individual transportation (0~points) & Public transportation  (1~point) & Active transportation (2~points) 
			\\ \midrule 
			Airplane&Bus&Bicycle\\
			Car&Cable car&Walking\\
			E-Scooter&Coach&\\
			Ferry&Rapid transit railway&\\
			Motorbike&Regional train&\\
			Other&Subway&\\
			&Train&\\
			&Tramway&
			\\ \bottomrule
		\end{tabular}
		\begin{tablenotes}
			\item Only actually detected means of transportation are displayed.
		\end{tablenotes}
	\end{threeparttable}
\end{table}
			
As we incentivized the maximization of these scores through both monetary incentives and informational nudges, they represent the economically sensible choice for the outcome variable $Y$. However, there are some caveats. First, since we did not provide a comprehensive list of means of transportation in our competition conditions, some discretion was necessary, as Table \ref{tab:scores} illustrates. Although airplanes and ferries are not strictly considered motorized individual transportation, no points were awarded for these trips, as doing so would clearly contradict the competition’s sustainability goals.

More importantly, no negative points were awarded. This means that sufficiency was not incentivized, as---in an extreme scenario---an airplane trip combined with two trips by foot could have yielded more points than staying at home. This approach was adopted for practical reasons: While we encouraged participants to enable permanent tracking, data protection concerns made it impossible to enforce this.\footnote{According to our competition conditions, eligibility for prizes required at least 30 days of tracking, each with a minimum of 6 hours recorded.} On the flip side, in conjunction with negative points, this would have opened the door for participants to game the competition by selectively suppressing the recording of penalized trips. As a result, in our implemented version of the competition, chances of winning were maximized by undertaking as many "sustainable trips" as possible, not necessarily by being CO\textsubscript{2} sufficient.\footnote{Since we were permitted to communicate the mobility competition under the name of HSLU, negative points could potentially have also clashed with the institution's guidelines, which do not promote absenteeism. For a further discussion on the efficacy of our treatment with respect to CO\textsubscript{2}, see Section \ref{discussion}.}

The disconnect between sustainable mobility and our outcome variable (and thus performance indicator) is further pronounced by the fact we only applied three values (see Table \ref{tab:scores}). What prevented us from implementing a subtler measurement was the need for ease of communication, as well as the multidimensionality of sustainable mobility, including considerations such as space efficiency and health. 

Finally, note that our outcome variable relates to trips, not kilometers. This approach was intended to level the playing field for participants regardless of their commuting distance. Moreover, expecting structural changes, such as participants relocating, would have been illusory.	

\subsection{Experimental Setting}\label{setting}

\begin{figure}[H]
	\centering
	\footnotesize
	\singlespacing	
	\renewcommand{\arraystretch}{1.1}
	\begin{tikzpicture}[
	node distance=0.4cm and 0.4cm,
	startstop/.style={
		rectangle,
		rounded corners,
		minimum width=2cm,
		minimum height=1.8cm,
		text width=2.4cm,
		align=center,
		draw=black,
		font=\footnotesize
	},
	arrow/.style={->,>=stealth}
	]
	
	% Nodes
	\node (start) [startstop] {\textbf{Step 1}: \\ Recruitment};
	\node (treatment) [startstop, right=of start] {\textbf{Step 2}: \\ Treatment allocation};
	\node (tracking) [startstop, right=of treatment, text width=5.5cm] {\textbf{Step 3}: \\ Tracking \\ (March to April 2024)};
	\node (winners) [startstop, right=of tracking] {\textbf{Step 5}: \\ Determination of winners; mobility report};
	%	\node (report) [startstop, right=of winners] {5. Mobility report};
	
	% Arrows
	\draw [arrow] (start) -- (treatment);
	\draw [arrow] (treatment) -- (tracking);
	\draw [arrow] (tracking) -- (winners);
	%	\draw [arrow] (winners) -- (report);
	
	% Communications
	\node (comm2) [above=2cm of tracking.north, xshift=-0.1cm, anchor=south, rotate=90, font=\footnotesize, text width=3cm] {Communication 2 (2024-03-28)};
	\node (comm1) [above=2cm of tracking.north,anchor=south, rotate=90, font=\footnotesize, text width=3cm, yshift=1.8cm] {Communication 1 (2024-03-07)};
	\node (comm3) [above=2cm of tracking.north, anchor=south, rotate=90, font=\footnotesize, text width=3cm, yshift=-1.5cm] {Communication 3 (2024-04-17)};
	
	% Downward arrows (adjusted for vertical placement)
	\draw [arrow] (comm1.south) -- ++(0,-2cm);
	\draw [arrow] (comm2.south) -- ++(0,-2cm);
	\draw [arrow] (comm3.south) -- ++(0,-2cm);
	
	% Draw brace
	\draw [decorate,decoration={brace,amplitude=10pt,raise=-10pt},yshift=0.5cm]
	(comm1.north east) -- ($(comm3.north east) + (0.9cm,0)$) node [black,midway,yshift=16pt] {\footnotesize \parbox{6cm}{\centering \textbf{Step 4}: \\ Interim communication}};
	
\end{tikzpicture}
	\caption{Process flow of the mobility competition}	\label{fig:process}
\end{figure}
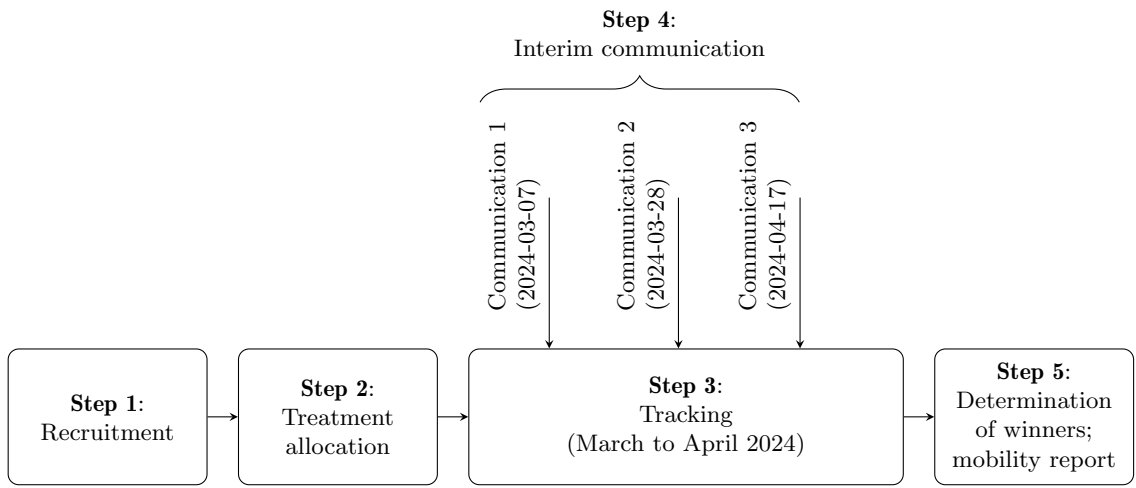

Our experiment, framed as a competition, followed the process outlined in Figure \ref{fig:process}. The competition was promoted as the "HSLU Mobility Challenge 2024" and took place in March and April 2024.\footnote{See \href{https://www.hslu.ch/de-ch/wirtschaft/institute/itm/mobilitaet/hslumobilitychallenge/}{https://www.hslu.ch/de-ch/wirtschaft/institute/itm/mobilitaet/hslumobilitychallenge} (accessed: 2024-08-06).\label{footnote:webpage}} In this subsection, we detail Steps 1 through 5.

\begin{enumerate}[leftmargin=*]

\item [\textbf{Step 1}]\label{step1}
The Lucerne University of Applied Sciences and Arts (HSLU) in Switzerland is a higher-education institution with 8,300 students, over 5,200 continuing-education participants, and around 1,900 employees. Participation in the challenge was open exclusively to HSLU students and employees. As our treatment structure ensured a randomized controlled trial, selection bias, discussed in Section \ref{discussion}, was of secondary importance, allowing us to adopt an aggressive promotion strategy. The competition was featured in a series of newsletters in late 2023, which served as stepping stones to our dedicated web page (see Footnote \ref{footnote:webpage}). Furthermore, physical flyers (alongside chocolate bars) were distributed at the main quarters of all six departments.
In addition to the promise of the total prize sum of CHF\,3,150 (see Step \hyperref[step3]{3}), we featured an additional prize draw, including a bicycle (valued CHF\,800) and four public-transportation vouchers (each valued CHF\,100) among participants who recorded data by the first day of the challenge. Preliminary registration took place on our web page. Agreeing to the competition conditions and acknowledging data-protection provisions occurred in a multi-step procedure via e-mail on February 27\textsuperscript{th} 2024. For details, see Excerpt \ref{ex:start} in Appendix \ref{app:excerpts}. 

A total of 264 participants joined the mobility challenge, with 200 achieving positive scores. The number of tracked participants per day is shown in Figure \ref{Fig:Timeseries_persons} in Appendix \ref{app:figures}. 

\item[\textbf{Step 2}]\label{step2}

The allocation to different groups based on incentives and communication strategies was balanced according to initial registrations (see Step~\hyperref[step4]{4} for detailed descriptions of the treatments). However, considering actual participation (after withdrawals) and respecting data subject rights (including the deletion of personal data upon request), the number of observations dropped to 195, resulting in the allocation depicted in Figure \ref{fig:groups}, which represents our 3$\times$3 factorial design.

%\vspace{-1em}
%\paragraph{Incentives} 

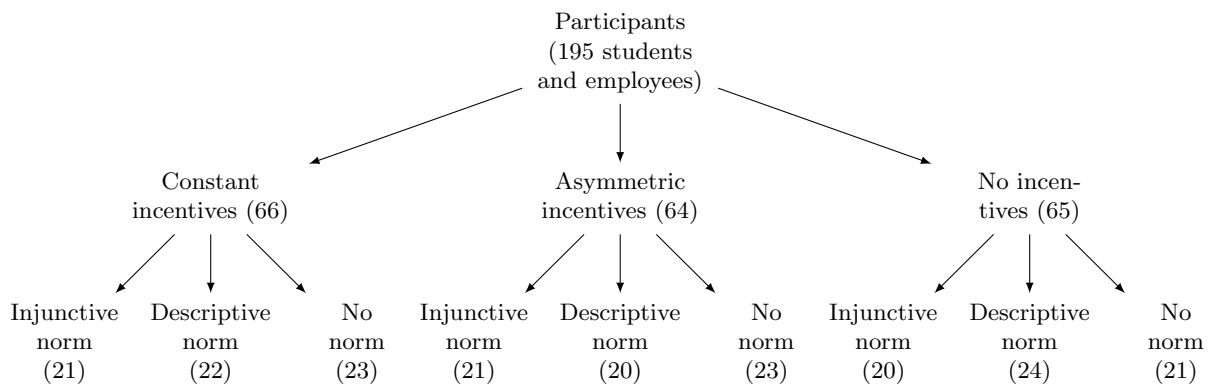
\begin{figure}[H]
	\begingroup
	\linespread{1}\selectfont 
		\begin{tikzpicture}[
	sibling distance=8em,
	level distance=8em,
	every node/.style={font=\footnotesize, align=center, text width=6em}, % Adjust text width as necessary
	edge from parent/.style={draw,-latex},
	level 1/.style={sibling distance=14em, level distance=5em}, % Adjust sibling distance for final nodes
	level 2/.style={sibling distance=5em, level distance=5em} % Adjust sibling distance for final nodes
	]
	\node {Participants (195 students and employees)}
	child {node {Constant incentives (66)}
		child {node {Injunctive \\ norm\\(21)}}
		child {node {Descriptive \\ norm\\(22)}}
		child {node {No \\ norm\\(23)}}
	}
	child {node {Asymmetric incentives (64)}
		child {node {Injunctive \\ norm\\(21)}}
		child {node {Descriptive \\ norm\\(20)}}
		child {node {No \\ norm\\(23)}}
	}
	child {node {No incentives (65)}
		child {node {Injunctive \\ norm\\(20)}}
		child {node {Descriptive \\ norm\\(24)}}
		child {node {No \\ norm\\(21)}}
	};
\end{tikzpicture}
	\endgroup
	\caption{Participant number by treatment}\label{fig:groups}
\end{figure}

\item[\textbf{Step 3}]\label{step3}

In mobility studies, the collection of data using tracking on smartphone apps has gained attention in recent years, particularly in studies on travel behavior, transportation safety, and transportation efficiency \citep{sadeghian2021review}. The technology we used, provided by MotionTag, is among the many that enhance traditional survey methods by employing online connectivity, geolocation, and smartphone applications \citep{bonnel2018transport}. It records the routes taken by participants and automatically detects their means of transportation, allowing stages and trips to be considered in addition to distances and travel times. Stages involve a single means of transportation, while trips consist of one or several stages and begin after stopovers of at least 20 minutes.\footnote{Note that MotionTag's definition of trips diverges from that of the Swiss Microcensus \citep{bfs_are2023}, which considers trips to be concluded only upon reaching a destination, such as work or shopping, where a specific purpose or activity takes place. As a result, our recorded average of 3.94 trips per day and participant exceeds the 2.8 reported by the Microcensus. However, the fact that our measured average trip duration of 33.28 minutes also exceeds the 29.2 minutes reported by the Microcensus suggests that we observe an above-average active sample—even before accounting for potentially unrecorded trips.}

As described in Subsection \ref{scores}, our outcome variable relates to the main means of transportation for a trip, which refers to the means of transportation used in the stage with the greatest distance. Although MotionTag's means detection is considered highly accurate---\cite{molloy2020national} concluded that for both major operating systems, it exceeded 92\% accuracy as early as 2019---some misassignments do occur. While our study participants could overwrite the detected modes using MotionTag's interface, we relied on the automatically detected modes to assign points and measure our outcome variable in order to discourage cheating and avoid biased results, as explicitly stated in our competition conditions.

\item[\textbf{Step 4}]\label{step4} 
In the course of our experiment, we sent three instances of interim communication via e-mail. Together with our competition conditions, summarized in the invitation e-mail (see Excerpt \ref{ex:start} in Appendix \ref{app:excerpts}), these messages exemplify our treatments regarding incentives and communication. 
	
%\vspace{-1em}
\paragraph{Incentives}
 
Rewarding well-performing participants can increase motivation to change behavior in the desired direction \citep{yen2019gamification}. Good performances in the challenge were incentivized with thematically appropriate prizes. Since the aim of our study is to measure different effects related to the incentive scheme rather than the prize volume, the expected profit (assuming identical mobility behavior) was the same in all groups. The sponsored prize amount of CHF\,3150 (roughly US\$\,3600) was equally divided among the participants of the three schemes: Each group had the same three prizes available---mobility vouchers valued at CHF\,600, CHF\,300, and CHF\,150---while the chances of winning depended on the scores. 

Under the \emph{Constant incentives} treatment, each active participant $i$ in the group of 66 participants had a chance to win a voucher, with the probability of winning $P_i$ proportional to their score $S_i$.\footnote{See Excerpt \ref{ex:constant_injunctive} in Appendix \ref{app:excerpts}.} Specifically, $P_i=S_i/\sum_{j=1}^{66}S_j$.  

Under the \emph{Asymmetric incentives} treatment, only the top three participants in the group of 64 participants received prizes; there was no prize draw.\footnote{See Excerpt \ref{ex:asymmetric_no} in Appendix \ref{app:excerpts}.} Depending on the type of communication and compared to the \emph{Constant incentives} treatment, this could have more strongly incentivized above-average performers to accumulate points in the final stretch of the experiment, potentially resulting in a "final sprint." Conversely, below-average participants may have been discouraged even more.

\emph{No incentives} describes the 65 participants without an incentive treatment.\footnote{See Excerpt \ref{ex:no_descriptive} in Appendix \ref{app:excerpts}.} In this control group, winners were drawn randomly. Thus, unlike in the other two groups, the probability of winning was independent of scores and, thus, mobility behavior.  
	
%\vspace{-1em}
\paragraph{Communication} 

Depending on how information is framed, communication can influence (mobility) behavior, extending beyond the rational-choice framework of mainstream economics. In our experiment, we categorized communication treatments according to a distinction typically made in the psychological literature, which refers to social norms \citetext{\citealp{cialdini1990focus}; \citealp{cialdini2003crafting}; \citealp{raux2021mobility}}. In contrast to personal norms, which are often aligned with political values (e.g., relating to ecological aspects), social norms are seen as perceived obligations for appropriate behavior within a social group, reflecting collective expectations. 

In our interim communication during the experiment, we applied two types of social norms in the context of transmitting temporary scores (see also Figure \ref{fig:groups}).

Under the \emph{Injunctive norms} treatment, 62 participants received information supplemented with reinforcement based on their current score, such as "Keep it up!" or "You can do better!"\footnote{See Excerpt \ref{ex:constant_injunctive} in Appendix \ref{app:excerpts}.} These messages convey how behavior should be in a given situation.

Under the \emph{Descriptive norms} treatment, 66 participants were provided with the mean value of their group's temporary score in addition to their individual score (where "group" refers to the incentives-related treatment).\footnote{See Excerpt \ref{ex:no_descriptive} in Appendix \ref{app:excerpts}.} Descriptive norms highlight the actual implementation of expected behavior, potentially reinforcing social expectations—a fact well understood by the advertising industry (think of "80\% of customers choose this subscription"). However, descriptive norms may also have adverse effects, given the (strong) assumption that above-average-performing individuals could also bend towards the social norm, inducing them to lessen their effort. 

Finally, \emph{No norm} refers to the 67 participants in the control group regarding the communication treatment.\footnote{See Excerpt \ref{ex:asymmetric_no} in Appendix \ref{app:excerpts}.} In the course of our interim communication, these participants were only communicated their individual current score, alongside procedural information (such as reminders) which was provided for all participants of the experiment.

As alluded to when describing the \emph{Asymmetric incentives} treatment, note that specific combinations of the incentives and communication treatments allow for additional participant strategies. For this reason, we included interaction terms (such as \emph{Asymmetric incentives} $\times$ \emph{Descriptive norm}) in our statistical analysis, which we present in the next section.

\item[\textbf{Step 5}]\label{step5}
Shortly after the experiment, on May 5\textsuperscript{th}, 2024, the lottery winners (under the \emph{Constant incentives} treatment and under \emph{No treatment}) and the participants with the highest scores (under the \emph{Asymmetric incentives} treatment) were awarded their prizes. 

All participants also received semi-automatically produced "mobility reports," which included statistics on their use of selected means of transportation and comparisons of their estimated individual CO\textsubscript{2} emissions with those of all participants and the general Swiss population (see Figures \ref{fig:kilometer} and \ref{fig:CO2} in Excerpt \ref{ex:mobility_report} in Appendix \ref{app:excerpts}). The mobility reports had a dual purpose: they were designed to encourage participants to reflect on their mobility behavior beyond the competition, reinforced by additional behavioral suggestions, and they served as a recruitment tool and a means to boost intrinsic motivation through emotionalization and gamification by announcing them at the competition's outset. 

\end{enumerate}

\section{Results}\label{results}

\subsection{Descriptive results}

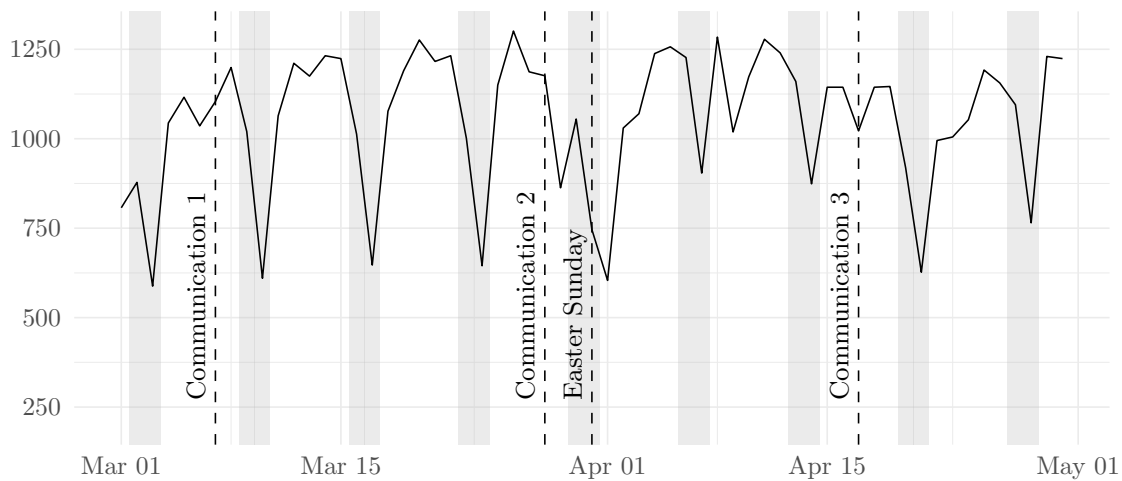
\begin{figure}[H] 
\begin{tikzpicture}[x=1pt,y=1pt]
	\definecolor{fillColor}{RGB}{255,255,255}
	\path[use as bounding box,fill=fillColor,fill opacity=0.00] (0,0) rectangle (433.62,216.81);
	\begin{scope}
		\path[clip] ( 40.51, 30.69) rectangle (428.12,194.15);
		\definecolor{drawColor}{gray}{0.92}
		
		\path[draw=drawColor,line width= 0.3pt,line join=round] ( 40.51, 61.74) --
		(428.12, 61.74);
		
		\path[draw=drawColor,line width= 0.3pt,line join=round] ( 40.51, 95.48) --
		(428.12, 95.48);
		
		\path[draw=drawColor,line width= 0.3pt,line join=round] ( 40.51,129.22) --
		(428.12,129.22);
		
		\path[draw=drawColor,line width= 0.3pt,line join=round] ( 40.51,162.97) --
		(428.12,162.97);
		
		\path[draw=drawColor,line width= 0.3pt,line join=round] ( 99.24, 30.69) --
		( 99.24,194.15);
		
		\path[draw=drawColor,line width= 0.3pt,line join=round] (190.27, 30.69) --
		(190.27,194.15);
		
		\path[draw=drawColor,line width= 0.3pt,line join=round] (281.30, 30.69) --
		(281.30,194.15);
		
		\path[draw=drawColor,line width= 0.3pt,line join=round] (369.39, 30.69) --
		(369.39,194.15);
		
		\path[draw=drawColor,line width= 0.6pt,line join=round] ( 40.51, 44.86) --
		(428.12, 44.86);
		
		\path[draw=drawColor,line width= 0.6pt,line join=round] ( 40.51, 78.61) --
		(428.12, 78.61);
		
		\path[draw=drawColor,line width= 0.6pt,line join=round] ( 40.51,112.35) --
		(428.12,112.35);
		
		\path[draw=drawColor,line width= 0.6pt,line join=round] ( 40.51,146.10) --
		(428.12,146.10);
		
		\path[draw=drawColor,line width= 0.6pt,line join=round] ( 40.51,179.84) --
		(428.12,179.84);
		
		\path[draw=drawColor,line width= 0.6pt,line join=round] ( 58.13, 30.69) --
		( 58.13,194.15);
		
		\path[draw=drawColor,line width= 0.6pt,line join=round] (140.35, 30.69) --
		(140.35,194.15);
		
		\path[draw=drawColor,line width= 0.6pt,line join=round] (240.19, 30.69) --
		(240.19,194.15);
		
		\path[draw=drawColor,line width= 0.6pt,line join=round] (322.41, 30.69) --
		(322.41,194.15);
		
		\path[draw=drawColor,line width= 0.6pt,line join=round] (416.37, 30.69) --
		(416.37,194.15);
		\definecolor{fillColor}{RGB}{190,190,190}
		
		\path[fill=fillColor,fill opacity=0.30] ( 61.07, 30.69) rectangle ( 66.94,194.15);
		
		\path[fill=fillColor,fill opacity=0.30] ( 66.94, 30.69) rectangle ( 72.81,194.15);
		
		\path[fill=fillColor,fill opacity=0.30] (102.18, 30.69) rectangle (108.05,194.15);
		
		\path[fill=fillColor,fill opacity=0.30] (108.05, 30.69) rectangle (113.92,194.15);
		
		\path[fill=fillColor,fill opacity=0.30] (143.29, 30.69) rectangle (149.16,194.15);
		
		\path[fill=fillColor,fill opacity=0.30] (149.16, 30.69) rectangle (155.03,194.15);
		
		\path[fill=fillColor,fill opacity=0.30] (184.40, 30.69) rectangle (190.27,194.15);
		
		\path[fill=fillColor,fill opacity=0.30] (190.27, 30.69) rectangle (196.14,194.15);
		
		\path[fill=fillColor,fill opacity=0.30] (225.51, 30.69) rectangle (231.38,194.15);
		
		\path[fill=fillColor,fill opacity=0.30] (231.38, 30.69) rectangle (237.25,194.15);
		
		\path[fill=fillColor,fill opacity=0.30] (266.62, 30.69) rectangle (272.49,194.15);
		
		\path[fill=fillColor,fill opacity=0.30] (272.49, 30.69) rectangle (278.36,194.15);
		
		\path[fill=fillColor,fill opacity=0.30] (307.73, 30.69) rectangle (313.60,194.15);
		
		\path[fill=fillColor,fill opacity=0.30] (313.60, 30.69) rectangle (319.47,194.15);
		
		\path[fill=fillColor,fill opacity=0.30] (348.84, 30.69) rectangle (354.71,194.15);
		
		\path[fill=fillColor,fill opacity=0.30] (354.71, 30.69) rectangle (360.58,194.15);
		
		\path[fill=fillColor,fill opacity=0.30] (389.95, 30.69) rectangle (395.82,194.15);
		
		\path[fill=fillColor,fill opacity=0.30] (395.82, 30.69) rectangle (401.69,194.15);
		\definecolor{drawColor}{RGB}{0,0,0}
		
		\path[draw=drawColor,line width= 0.6pt,line join=round] ( 58.13,120.05) --
		( 64.00,129.63) --
		( 69.87, 90.49) --
		( 75.75,152.03) --
		( 81.62,161.75) --
		( 87.49,150.95) --
		( 93.37,160.13) --
		( 99.24,172.96) --
		(105.11,148.79) --
		(110.98, 93.46) --
		(116.86,154.73) --
		(122.73,174.57) --
		(128.60,169.72) --
		(134.48,177.41) --
		(140.35,176.33) --
		(146.22,147.85) --
		(152.09, 98.45) --
		(157.97,156.49) --
		(163.84,171.74) --
		(169.71,183.35) --
		(175.59,175.25) --
		(181.46,177.41) --
		(187.33,146.23) --
		(193.20, 98.18) --
		(199.08,166.34) --
		(204.95,186.72) --
		(210.82,171.34) --
		(216.70,169.85) --
		(222.57,127.60) --
		(228.44,153.52) --
		(234.32,111.81) --
		(240.19, 92.65) --
		(246.06,150.14) --
		(251.93,155.54) --
		(257.81,178.22) --
		(263.68,180.78) --
		(269.55,176.73) --
		(275.43,133.14) --
		(281.30,184.43) --
		(287.17,148.66) --
		(293.04,169.45) --
		(298.92,183.62) --
		(304.79,178.49) --
		(310.66,167.69) --
		(316.54,129.09) --
		(322.41,165.53) --
		(328.28,165.53) --
		(334.15,149.06) --
		(340.03,165.53) --
		(345.90,165.80) --
		(351.77,135.16) --
		(357.65, 95.75) --
		(363.52,145.42) --
		(369.39,146.77) --
		(375.26,153.25) --
		(381.14,172.01) --
		(387.01,167.15) --
		(392.88,158.92) --
		(398.76,114.38) --
		(404.63,177.14) --
		(410.50,176.33);
		
		\path[draw=drawColor,line width= 0.6pt,dash pattern=on 4pt off 4pt ,line join=round] ( 93.37, 30.69) -- ( 93.37,194.15);
		
		\path[draw=drawColor,line width= 0.6pt,dash pattern=on 4pt off 4pt ,line join=round] (216.70, 30.69) -- (216.70,194.15);
		
		\path[draw=drawColor,line width= 0.6pt,dash pattern=on 4pt off 4pt ,line join=round] (334.15, 30.69) -- (334.15,194.15);
		
		\path[draw=drawColor,line width= 0.6pt,dash pattern=on 4pt off 4pt ,line join=round] (234.32, 30.69) -- (234.32,194.15);
		
		\node[text=drawColor,rotate= 90.00,anchor=base west,inner sep=0pt, outer sep=0pt, scale=  1.10] at ( 89.56, 47.53) {\footnotesize Communication 1};
		
		\node[text=drawColor,rotate= 90.00,anchor=base west,inner sep=0pt, outer sep=0pt, scale=  1.10] at (212.89, 47.53) {\footnotesize Communication 2};
		
		\node[text=drawColor,rotate= 90.00,anchor=base west,inner sep=0pt, outer sep=0pt, scale=  1.10] at (330.35, 47.53) {\footnotesize Communication 3};
		
		\node[text=drawColor,rotate= 90.00,anchor=base west,inner sep=0pt, outer sep=0pt, scale=  1.10] at (230.51, 47.53) {\footnotesize Easter Sunday};
	\end{scope}
	\begin{scope}
		\path[clip] (  0.00,  0.00) rectangle (433.62,216.81);
		\definecolor{drawColor}{gray}{0.30}
		
		\node[text=drawColor,anchor=base east,inner sep=0pt, outer sep=0pt, scale=  0.88] at ( 35.56, 41.83) {250};
		
		\node[text=drawColor,anchor=base east,inner sep=0pt, outer sep=0pt, scale=  0.88] at ( 35.56, 75.58) {500};
		
		\node[text=drawColor,anchor=base east,inner sep=0pt, outer sep=0pt, scale=  0.88] at ( 35.56,109.32) {750};
		
		\node[text=drawColor,anchor=base east,inner sep=0pt, outer sep=0pt, scale=  0.88] at ( 35.56,143.06) {1000};
		
		\node[text=drawColor,anchor=base east,inner sep=0pt, outer sep=0pt, scale=  0.88] at ( 35.56,176.81) {1250};
	\end{scope}
	\begin{scope}
		\path[clip] (  0.00,  0.00) rectangle (433.62,216.81);
		\definecolor{drawColor}{gray}{0.30}
		
		\node[text=drawColor,anchor=base,inner sep=0pt, outer sep=0pt, scale=  0.88] at ( 58.13, 19.68) {Mar 01};
		
		\node[text=drawColor,anchor=base,inner sep=0pt, outer sep=0pt, scale=  0.88] at (140.35, 19.68) {Mar 15};
		
		\node[text=drawColor,anchor=base,inner sep=0pt, outer sep=0pt, scale=  0.88] at (240.19, 19.68) {Apr 01};
		
		\node[text=drawColor,anchor=base,inner sep=0pt, outer sep=0pt, scale=  0.88] at (322.41, 19.68) {Apr 15};
		
		\node[text=drawColor,anchor=base,inner sep=0pt, outer sep=0pt, scale=  0.88] at (416.37, 19.68) {May 01};
	\end{scope}
\end{tikzpicture}
	\centering \caption{Time series of aggregated daily points (March and April 2024; weekends shaded)} \label{Fig:timeseries_points}
\end{figure}

As we show in Figure \ref{Fig:timeseries_points}, participants cumulatively accrued between 500 and nearly 1500 points per day. This corresponds to individual daily averages of between 2.6 and 7.7 points. Furthermore, the figure illustrates weekly cycles, with the least point accumulation occurring on Sundays.\footnote{See Figures \ref{Fig:Timeseries_point_group} in Appendix \ref{app:figures} for an analogous plot grouped by incentive scheme, and Figure \ref{Fig:Timeseries_CO2} in Appendix \ref{app:figures} with CO$_2$ emissions as the dependent variable.}

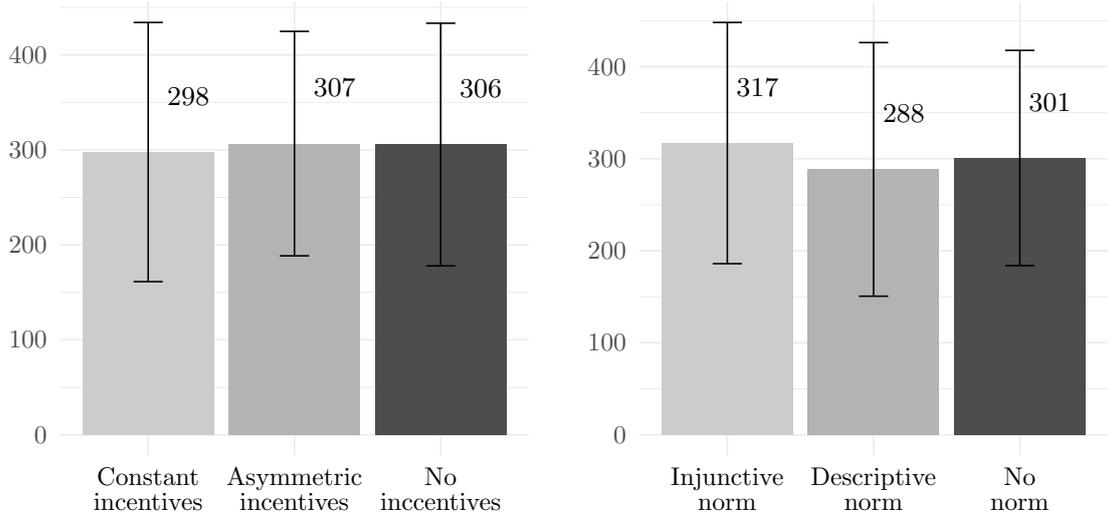
\begin{figure}
		\begin{tikzpicture}[x=1pt,y=1pt]
	\definecolor{fillColor}{RGB}{255,255,255}
	\path[use as bounding box,fill=fillColor,fill opacity=0.00] (0,0) rectangle (433.62,216.81);
	\begin{scope}
		\path[clip] ( 36.11, 40.19) rectangle (211.31,211.31);
		\definecolor{drawColor}{gray}{0.92}
		
		\path[draw=drawColor,line width= 0.3pt,line join=round] ( 36.11, 65.88) --
		(211.31, 65.88);
		
		\path[draw=drawColor,line width= 0.3pt,line join=round] ( 36.11,101.72) --
		(211.31,101.72);
		
		\path[draw=drawColor,line width= 0.3pt,line join=round] ( 36.11,137.55) --
		(211.31,137.55);
		
		\path[draw=drawColor,line width= 0.3pt,line join=round] ( 36.11,173.38) --
		(211.31,173.38);
		
		\path[draw=drawColor,line width= 0.3pt,line join=round] ( 36.11,209.21) --
		(211.31,209.21);
		
		\path[draw=drawColor,line width= 0.6pt,line join=round] ( 36.11, 47.97) --
		(211.31, 47.97);
		
		\path[draw=drawColor,line width= 0.6pt,line join=round] ( 36.11, 83.80) --
		(211.31, 83.80);
		
		\path[draw=drawColor,line width= 0.6pt,line join=round] ( 36.11,119.63) --
		(211.31,119.63);
		
		\path[draw=drawColor,line width= 0.6pt,line join=round] ( 36.11,155.46) --
		(211.31,155.46);
		
		\path[draw=drawColor,line width= 0.6pt,line join=round] ( 36.11,191.30) --
		(211.31,191.30);
		
		\path[draw=drawColor,line width= 0.6pt,line join=round] ( 68.96, 40.19) --
		( 68.96,211.31);
		
		\path[draw=drawColor,line width= 0.6pt,line join=round] (123.71, 40.19) --
		(123.71,211.31);
		
		\path[draw=drawColor,line width= 0.6pt,line join=round] (178.46, 40.19) --
		(178.46,211.31);
		\definecolor{fillColor}{gray}{0.80}
		
		\path[fill=fillColor] ( 44.32, 47.97) rectangle ( 93.60,154.64);
		\definecolor{fillColor}{gray}{0.70}
		
		\path[fill=fillColor] ( 99.07, 47.97) rectangle (148.35,157.83);
		\definecolor{fillColor}{gray}{0.30}
		
		\path[fill=fillColor] (153.82, 47.97) rectangle (203.10,157.47);
		\definecolor{drawColor}{RGB}{0,0,0}
		
		\path[draw=drawColor,line width= 0.6pt,line join=round] ( 63.49,203.53) --
		( 74.44,203.53);
		
		\path[draw=drawColor,line width= 0.6pt,line join=round] ( 68.96,203.53) --
		( 68.96,105.75);
		
		\path[draw=drawColor,line width= 0.6pt,line join=round] ( 63.49,105.75) --
		( 74.44,105.75);
		
		\path[draw=drawColor,line width= 0.6pt,line join=round] (118.24,200.17) --
		(129.19,200.17);
		
		\path[draw=drawColor,line width= 0.6pt,line join=round] (123.71,200.17) --
		(123.71,115.49);
		
		\path[draw=drawColor,line width= 0.6pt,line join=round] (118.24,115.49) --
		(129.19,115.49);
		
		\path[draw=drawColor,line width= 0.6pt,line join=round] (172.99,203.24) --
		(183.94,203.24);
		
		\path[draw=drawColor,line width= 0.6pt,line join=round] (178.46,203.24) --
		(178.46,111.70);
		
		\path[draw=drawColor,line width= 0.6pt,line join=round] (172.99,111.70) --
		(183.94,111.70);
		
		\node[text=drawColor,anchor=base west,inner sep=0pt, outer sep=0pt, scale=  1.14] at ( 76.11,172.51) {\footnotesize 298};
		
		\node[text=drawColor,anchor=base west,inner sep=0pt, outer sep=0pt, scale=  1.14] at (130.86,175.70) {\footnotesize 307};
		
		\node[text=drawColor,anchor=base west,inner sep=0pt, outer sep=0pt, scale=  1.14] at (185.61,175.34) {\footnotesize 306};
	\end{scope}
	\begin{scope}
		\path[clip] (  0.00,  0.00) rectangle (433.62,216.81);
		\definecolor{drawColor}{gray}{0.30}
		
		\node[text=drawColor,anchor=base east,inner sep=0pt, outer sep=0pt, scale=  0.88] at ( 31.16, 44.94) {0};
		
		\node[text=drawColor,anchor=base east,inner sep=0pt, outer sep=0pt, scale=  0.88] at ( 31.16, 80.77) {100};
		
		\node[text=drawColor,anchor=base east,inner sep=0pt, outer sep=0pt, scale=  0.88] at ( 31.16,116.60) {200};
		
		\node[text=drawColor,anchor=base east,inner sep=0pt, outer sep=0pt, scale=  0.88] at ( 31.16,152.43) {300};
		
		\node[text=drawColor,anchor=base east,inner sep=0pt, outer sep=0pt, scale=  0.88] at ( 31.16,188.27) {400};
	\end{scope}
	\begin{scope}
		\path[clip] (  0.00,  0.00) rectangle (433.62,216.81);
		\definecolor{drawColor}{gray}{0.30}
		
		\node[text=black,anchor=base,inner sep=0pt, outer sep=0pt, scale=  0.88] at ( 68.96, 29.18) {Constant};
		
		\node[text=black,anchor=base,inner sep=0pt, outer sep=0pt, scale=  0.88] at ( 68.96, 19.68) {incentives};
		
		\node[text=black,anchor=base,inner sep=0pt, outer sep=0pt, scale=  0.88] at (123.71, 29.18) {Asymmetric};
		
		\node[text=black,anchor=base,inner sep=0pt, outer sep=0pt, scale=  0.88] at (123.71, 19.68) {incentives};
		
		\node[text=black,anchor=base,inner sep=0pt, outer sep=0pt, scale=  0.88] at (178.46, 29.18) {No};
		
		\node[text=black,anchor=base,inner sep=0pt, outer sep=0pt, scale=  0.88] at (178.46, 19.68) {inccentives};
	\end{scope}
	\begin{scope}
		\path[clip] (252.92, 40.19) rectangle (428.12,211.31);
		\definecolor{drawColor}{gray}{0.92}
		
		\path[draw=drawColor,line width= 0.3pt,line join=round] (252.92, 65.33) --
		(428.12, 65.33);
		
		\path[draw=drawColor,line width= 0.3pt,line join=round] (252.92,100.05) --
		(428.12,100.05);
		
		\path[draw=drawColor,line width= 0.3pt,line join=round] (252.92,134.77) --
		(428.12,134.77);
		
		\path[draw=drawColor,line width= 0.3pt,line join=round] (252.92,169.49) --
		(428.12,169.49);
		
		\path[draw=drawColor,line width= 0.3pt,line join=round] (252.92,204.21) --
		(428.12,204.21);
		
		\path[draw=drawColor,line width= 0.6pt,line join=round] (252.92, 47.97) --
		(428.12, 47.97);
		
		\path[draw=drawColor,line width= 0.6pt,line join=round] (252.92, 82.69) --
		(428.12, 82.69);
		
		\path[draw=drawColor,line width= 0.6pt,line join=round] (252.92,117.41) --
		(428.12,117.41);
		
		\path[draw=drawColor,line width= 0.6pt,line join=round] (252.92,152.13) --
		(428.12,152.13);
		
		\path[draw=drawColor,line width= 0.6pt,line join=round] (252.92,186.85) --
		(428.12,186.85);
		
		\path[draw=drawColor,line width= 0.6pt,line join=round] (285.77, 40.19) --
		(285.77,211.31);
		
		\path[draw=drawColor,line width= 0.6pt,line join=round] (340.52, 40.19) --
		(340.52,211.31);
		
		\path[draw=drawColor,line width= 0.6pt,line join=round] (395.27, 40.19) --
		(395.27,211.31);
		\definecolor{fillColor}{gray}{0.80}
		
		\path[fill=fillColor] (261.13, 47.97) rectangle (310.41,158.03);
		\definecolor{fillColor}{gray}{0.70}
		
		\path[fill=fillColor] (315.88, 47.97) rectangle (365.16,148.10);
		\definecolor{fillColor}{gray}{0.30}
		
		\path[fill=fillColor] (370.63, 47.97) rectangle (419.91,152.40);
		\definecolor{drawColor}{RGB}{0,0,0}
		
		\path[draw=drawColor,line width= 0.6pt,line join=round] (280.30,203.53) --
		(291.25,203.53);
		
		\path[draw=drawColor,line width= 0.6pt,line join=round] (285.77,203.53) --
		(285.77,112.53);
		
		\path[draw=drawColor,line width= 0.6pt,line join=round] (280.30,112.53) --
		(291.25,112.53);
		
		\path[draw=drawColor,line width= 0.6pt,line join=round] (335.05,195.94) --
		(346.00,195.94);
		
		\path[draw=drawColor,line width= 0.6pt,line join=round] (340.52,195.94) --
		(340.52,100.26);
		
		\path[draw=drawColor,line width= 0.6pt,line join=round] (335.05,100.26) --
		(346.00,100.26);
		
		\path[draw=drawColor,line width= 0.6pt,line join=round] (389.80,193.00) --
		(400.75,193.00);
		
		\path[draw=drawColor,line width= 0.6pt,line join=round] (395.27,193.00) --
		(395.27,111.81);
		
		\path[draw=drawColor,line width= 0.6pt,line join=round] (389.80,111.81) --
		(400.75,111.81);
		
		\node[text=drawColor,anchor=base west,inner sep=0pt, outer sep=0pt, scale=  1.14] at (289.18,175.90) {\footnotesize 317};
		
		\node[text=drawColor,anchor=base west,inner sep=0pt, outer sep=0pt, scale=  1.14] at (343.93,165.97) {\footnotesize 288};
		
		\node[text=drawColor,anchor=base west,inner sep=0pt, outer sep=0pt, scale=  1.14] at (398.68,170.28) {\footnotesize 301};
	\end{scope}
	\begin{scope}
		\path[clip] (  0.00,  0.00) rectangle (433.62,216.81);
		\definecolor{drawColor}{gray}{0.30}
		
		\node[text=drawColor,anchor=base east,inner sep=0pt, outer sep=0pt, scale=  0.88] at (247.97, 44.94) {0};
		
		\node[text=drawColor,anchor=base east,inner sep=0pt, outer sep=0pt, scale=  0.88] at (247.97, 79.66) {100};
		
		\node[text=drawColor,anchor=base east,inner sep=0pt, outer sep=0pt, scale=  0.88] at (247.97,114.38) {200};
		
		\node[text=drawColor,anchor=base east,inner sep=0pt, outer sep=0pt, scale=  0.88] at (247.97,149.10) {300};
		
		\node[text=drawColor,anchor=base east,inner sep=0pt, outer sep=0pt, scale=  0.88] at (247.97,183.82) {400};
	\end{scope}
	\begin{scope}
		\path[clip] (  0.00,  0.00) rectangle (433.62,216.81);
		\definecolor{drawColor}{gray}{0.30}
		
		\node[text=black,anchor=base,inner sep=0pt, outer sep=0pt, scale=  0.88] at (285.77, 29.18) {Injunctive};
		
		\node[text=black,anchor=base,inner sep=0pt, outer sep=0pt, scale=  0.88] at (285.77, 19.68) {norm};
		
		\node[text=black,anchor=base,inner sep=0pt, outer sep=0pt, scale=  0.88] at (340.52, 29.18) {Descriptive};
		
		\node[text=black,anchor=base,inner sep=0pt, outer sep=0pt, scale=  0.88] at (340.52, 19.68) {norm};
		
		\node[text=black,anchor=base,inner sep=0pt, outer sep=0pt, scale=  0.88] at (395.27, 29.18) {No};
		
		\node[text=black,anchor=base,inner sep=0pt, outer sep=0pt, scale=  0.88] at (395.27, 19.68) {norm};
	\end{scope}
\end{tikzpicture}
		\centering \caption{Average scores (left: incentives treatments; right: communication treatments)} \label{Fig:barplot}
\end{figure}

 In Figure \ref{Fig:barplot}, we present average scores grouped by the incentive treatment (left panel) and the communication treatment (right panel), including standard deviations represented by error bars. Anticipating our causal results, the left-hand panel illustrates the absence of significant differences between participants receiving monetary incentives (\emph{Constant incentives} and \emph{Asymmetric incentives} treatments) and the control group (\emph{No incentives}). Individuals under the \emph{Injunctive norm} communication treatment have a slightly higher average score. Again, taking standard errors into account suggests that no significant causal treatment effects are to be expected regarding communication either.

\subsection{No Causal Effects on Sustainable Mobility as Measured}

We present our main causal estimations in Table \ref{tab:eff_full}. Employing the properties of our 3$\times$3 factorial design in a randomized controlled trial and in an attempt to maximize statistical power, we consider only either the incentives or the communication treatments in Regressions \hyperref[tab:eff_full]{(1)} and \hyperref[tab:eff_full]{(2)}, respectively.\footnote{Note that we include the control variable \emph{Student} in all of our regressions. This is to prevent omitted variable bias, as we did not consider whether a participant is a student or staff member in our factorial design, but we observe that, depending on the incentives/communication treatment combination, the proportion of students substantially varies between 47.8\% and 90\%.} In Regression \hyperref[tab:eff_full]{(3)}, we estimate the treatment effects regarding incentives and communication simultaneously. In Regression \hyperref[tab:eff_full]{(4)}, we include interaction effects to detect any potential synergistic treatment effects. However, as with the other regressions, all we can show is that our participants' mobility behavior, as measured by scores (our outcome variable), is stubbornly non-responsive to all our treatments, both regarding incentives and communication. While the values of R$^2$ naturally increase with the number of covariates, all our estimates exhibit negative values of adjusted R$^2$.\footnote{In a further attempt to increase the precision of our estimates, in the regressions shown in Table \ref{tab:effects_simple} in Appendix \ref{app:tables}, we simplify the model even further by defining binary \emph{Incentives} and \emph{Communication} treatments, which in both cases apply to all participants except the control groups. However, the negative values of adjusted R$^2$ persist.} 

\begin{table}[H]
	\centering
	\footnotesize
	\singlespacing	
	\renewcommand{\arraystretch}{1.1}
	\begin{threeparttable}	
	\caption{Regression of scores on incentives and communication}\label{tab:eff_full}
	\begin{tabular}{p{3cm}cccc}      
		\toprule
		& (1) & (2) & (3) & (4) \\ \midrule %% TinyTableHeader
		\multirow{2}{3cm}{(Intercept)}                                                               & \num{313.04}***                & \num{306.06}***                & \num{311.33}***                & \num{329.18}***                 \\
		& [\num{272.07}, \num{354.02}] & [\num{267.42}, \num{344.71}] & [\num{263.77}, \num{358.89}] & [\num{266.42}, \num{391.94}] \vspace{7pt} \\
		\multirow{2}{3cm}{Student}                                                              & \num{-10.95}                   & \num{-8.85}                    & \num{-9.78}                    & \num{-13.05}                    \\
		& [\num{-49.39}, \num{27.49}]  & [\num{-47.29}, \num{29.59}]  & [\num{-48.46}, \num{28.91}]  & [\num{-52.61}, \num{26.52}]  \vspace{7pt} \\
		\multirow{2}{3cm}{Constant incentives (CI)}                                      & \num{-13.23}                   &                                  & \num{-13.73}                   & \num{-34.76}                    \\
		& [\num{-57.91}, \num{31.46}]  &                                  & [\num{-58.48}, \num{31.03}]  & [\num{-112.37}, \num{42.86}] \vspace{7pt} \\
		\multirow{2}{3cm}{Asymmetric incentives (AI)}                                   & \num{0.96}                     &                                  & \num{-0.03}                    & \num{-25.29}                    \\
		& [\num{-43.99}, \num{45.90}]  &                                  & [\num{-45.08}, \num{45.03}]  & [\num{-103.35}, \num{52.78}] \vspace{7pt} \\
		\multirow{2}{3cm}{Injunctive norm (IN)}                                            &                                  & \num{16.67}                    & \num{16.65}                    & \num{-19.70}                    \\
		&                                  & [\num{-28.23}, \num{61.56}]  & [\num{-28.43}, \num{61.72}]  & [\num{-99.97}, \num{60.57}] \vspace{7pt}  \\
				\multirow{2}{3cm}{Descriptive norm (DN)}                                           &                                  & \num{-11.36}                   & \num{-11.40}                   & \num{-23.44}                    \\
		&                                  & [\num{-55.73}, \num{33.00}]  & [\num{-55.98}, \num{33.20}]  & [\num{-100.19}, \num{53.32}] \vspace{7pt} \\
		\multirow{2}{3cm}{CI × IN}     &                                  &                                  &                                  & \num{64.64}                     \\
		&                                  &                                  &                                  & [\num{-46.95}, \num{176.23}] \vspace{7pt} \\
		\multirow{2}{3cm}{CI × DN}    &                                  &                                  &                                  & \num{0.55}                      \\
		&                                  &                                  &                                  & [\num{-107.87}, \num{108.95}] \vspace{7pt} \\
		\multirow{2}{3cm}{AI × IN}  &                                  &                                  &                                  & \num{42.50}                     \\
&                                  &                                  &                                  & [\num{-69.50}, \num{154.50}] \vspace{7pt} \\
		\multirow{2}{3cm}{AI × DN} &                                  &                                  &                                  & \num{36.34}                     \\
		&                                  &                                  &                                  & [\num{-75.00}, \num{147.67}] \vspace{7pt} \\
		R$^2$                                                                        & \num{0.004}                     & \num{0.009}                     & \num{0.012}                     & \num{0.024}                     \vspace{7pt} \\
		Adjusted R$^2$                                      & \num{-0.012}                     & \num{-0.006}                     & \num{-0.014}                     & \num{-0.024}                     \\
		\bottomrule
	\end{tabular}
	\begin{tablenotes}[flushleft]
		\setlength{\itemsep}{2pt} % Adjust item separation
		\item Number of observations: 195.
		\item * p < 0.1, ** p < 0.05, *** p < 0.01.
	\end{tablenotes}
	\end{threeparttable}
\end{table} 

This main result remains in effect when adjusting our outcome variable to include only points accrued from either public transportation or active transportation. A rationale for doing so is the suggestion of heterogeneous price elasticities (or rather "prize elasticities") regarding these modes of transportation. However, the only significant insights from these additional regressions are that staff members gain more points through active transportation, while students gain more points through public transportation.\footnote{While there are obvious economic factors that explain why students choose public transportation more often, the reason for the higher scores for active transportation among staff members is less clear. We hypothesize that the selection of sustainability-aware participants into the experiment was potentially more pronounced among staff members.} Since these findings do not alter our overall conclusions, we omit the corresponding regression outputs from this paper. Furthermore, to avoid the risk of engaging in "p-hacking" \citep[see, e.g., ][]{simonsohn2014p} we refrain from conducting even more nuanced forms of post-hoc analyses. Our key finding, therefore, remains that neither our incentives treatments nor our communication treatments are capable of inducing sustainable mobility, as measured by our outcome variable.

Note that even in the case of Regression \hyperref[tab:eff_full]{(4)}, where we have nine variables (including treatments and controls), Cohen's $f^2$ for a well-powered test ($\beta = 0.2$) with a significance level of $\alpha = 0.05$ and our 195 observations is 0.08. This indicates that even relatively weak treatment effects would have been detectable. 

\subsection{Do We Produce a "Final Sprint"?}

Before drawing final conclusions, we consider a unique aspect of the monetary incentives tied to the information provided during our interim communications. Under the \emph{Asymmetric incentives} treatment, participants' knowledge of their relative performance compared to competitors can either enhance or diminish their extrinsic motivation. Specifically, when a participant is performing above average, maintaining their current pace results in a predictably higher probability of winning. This contrasts with the \emph{Constant incentives}  treatment (as well as with \emph{No incentives}), where the marginal contribution to the chances of winning remains constant, and with participants who do not receive information about their relative performance.

In the regression shown in Table \ref{tab:eff_top}, we adjust our outcome variable to consider only the points accrued after the final interim communication on April 17\textsuperscript{th}. 

%\begin{table}[H]
%	\centering 
%	\footnotesize
%	\singlespacing	
%	\begin{threeparttable}
%		\caption{Regression of scores after final communication on top-contender status and information about high chances of winning} 
%		\label{tab:eff_top} 
%		\begin{tabular}{p{4cm}c} 
%			\toprule 
%			\multirow{2}{4cm}{(Intercept)} & 60.72$^{***}$ \\ 
%			& [52.54, 68.89] \vspace{7pt}\\ 
%			\multirow{2}{4cm}{Student} & $-$6.85 \\ 
%			& [$-$16.08, 2.39] \vspace{7pt}\\ 
%			\multirow{2}{4cm}{Top contender} & 50.19$^{***}$ \\ 
%			& [39.68, 60.70] \vspace{7pt}\\ 
%			\multirow{2}{4cm}{Information} & 0.60 \\ 
%			& [$-$12.64, 13.85] \vspace{7pt}\\ 
%			\multirow{2}{2.5cm}{Top contender × Information} & 3.96 \\ 
%			& [$-$18.67, 26.58] \vspace{7pt}\\ 
%			R$^{2}$ & 0.383 \vspace{7pt}\\ 
%			Adjusted R$^{2}$ & 0.370\\ 
%			\bottomrule 
%		\end{tabular} 
%		\begin{tablenotes}[flushleft]
%			\setlength{\itemsep}{2pt} % Adjust item separation
%			\item Number of observations: 195.
%			\item Top contender = 1: Points before final communication within top 40\% in incentives category.
%			\item Information = 1: Asymmetric incentives \emph{and} Communication $\in$ \{Injunctive norm, Descriptive norm\}. 
%			\item * p < 0.1, ** p < 0.05, *** p < 0.01.
%		\end{tablenotes}
%	\end{threeparttable}
%\end{table} 

\begin{table}[H]
	\centering 
	\footnotesize
	\singlespacing	
	\begin{threeparttable}
		\caption{Regression of scores after final communication on top-contender status and information about high chances of winning} 
		\label{tab:eff_top} 
		\begin{tabular}{lc} 
			\toprule 
			Intercept & 60.72$^{***}$ [52.54, 68.89] \vspace{7pt}\\ 
			Student & $-$6.85 [$-$16.08, 2.39] \vspace{7pt}\\ 
			Top contender & 50.19$^{***}$ [39.68, 60.70] \vspace{7pt}\\ 
			Information & 0.60 [$-$12.64, 13.85] \vspace{7pt}\\ 
			Top contender × Information & 3.96 [$-$18.67, 26.58] \vspace{7pt}\\ 
			R$^{2}$ & 0.383 \vspace{7pt}\\ 
			Adjusted R$^{2}$ & 0.370\\ 
			\bottomrule 
		\end{tabular} 
		\begin{tablenotes}[flushleft]
			\setlength{\itemsep}{2pt} % Adjust item separation
			\item Number of observations: 195.
			\item Top contender = 1: Points before final communication within top 40\% in incentives category.
			\item Information = 1: Asymmetric incentives \emph{and} Communication $\in$ \{Injunctive norm, Descriptive norm\}. 
			\item * p < 0.1, ** p < 0.05, *** p < 0.01.
		\end{tablenotes}
	\end{threeparttable}
\end{table} 

Our focus is on the interaction between \emph{Top contender} (being within a group's top 40\% at the time of this final communication) and \emph{Information} (being treated with \emph{Asymmetric incentives} and simultaneously informed about relative performance, which occurs under both the \emph{Injunctive norm} and \emph{Descriptive norm} treatments).

Alas, as the coefficient's wide confidence interval of -18.67 to 26.58 illustrates, we do not observe even this supposedly strong incentive having an impact on behavior---and the significant coefficient of \emph{Top contender} is unsurprising, as a simple continuation of previous behavior may sufficiently explain it.

\section{Discussion}\label{discussion}

Before recapitulating our (null) results and discussing possible explanations in Section \ref{conclusion}, we briefly address two issues that should be considered when interpreting our findings, regardless of whether they suggest significant treatment effects or---as in our case---indicate that the treatments have no impact on sustainable mobility behavior.

First, recall that our outcome variable differs from sustainable mobility, as discussed in Subsection \ref{scores}. To assess the correlation between our scores and environmentally sustainable mobility, we calculate the sum of each participant's CO\textsubscript{2} emissions as an indicator of the latter.\footnote{We also made use of these CO\textsubscript{2} calculations in the context of our mobility reports; see Step \hyperref[step5]{5} in Subsection \ref{setting}.} Following, for instance, \cite{ohnmacht2020relationships} and \cite{balthasar2024effects}, we apply the Swiss "mobitool factors" \citep[see][]{mobitool2023}. These factors, employed on passenger kilometers by means of transportation in a product-sum manner, account for direct emissions from a vehicle's exhaust system, as well as emissions from vehicle maintenance and the construction and maintenance of infrastructure. In line with the design of our outcome variable, and as we exemplify in Table \ref{tab:mobitool} in Appendix \ref{app:tables}, the values active transportation are substantially smaller than those for public transportation and---in particular---motorized individual transportation. Using a simple linear regression, as illustrated in the scatterplot in Figure \ref{Fig:Scatter_Points_CO2} in Appendix \ref{app:figures}, we find that our scores are essentially unrelated\footnote{The R$^2$ of the simple regression is 0.00.} to this measure of sustainability. This lack of correlation stems from a key issue: while we correctly award points for choosing CO\textsubscript{2}-efficient transportation, participants who travel more frequently inevitably accumulate more points, regardless of the sustainability of their choices. Since the primary objective of the competition was to encourage the use of more sustainable transportation options rather than to reduce overall travel, this lack of association is less problematic than it might initially seem.

Second, it is clear that our study sample does not represent the total population (say, of Switzerland). Instead, it consists of a self-selected group of individuals who chose to participate in the mobility challenge, and it is plausible that they exhibit above-average motivation to change their mobility behavior. This does not, however, imply that the results of our experiment are biased. Rather, our estimated coefficients should be interpreted as \emph{conditional} average treatment effects; that is, conditional on joining the competition. On the other hand, this distinction may be somewhat academic, given the lack of significant effects in the first place.

\section{Conclusion}\label{conclusion}

To sum up, we conducted a randomized controlled trial in the form of a mobility competition to investigate whether particular types of competition formats (incentives treatments) or the ways in which we communicate with participants (communication treatments) succeed in improving mobility behavior, measured by the number of trips covered by means of active transportation and public transportation. The resulting insights could have helped in adapting future versions of mobility competitions, at the HSLU or elsewhere.

The basis for unearthing causal and meaningful results was laid. In addition to the random assignment of participants according to our 3$\times$3 factorial design, non-intrusive and state-of-the-art tracking allowed for precise measurement of trips by different means of transportation, our chosen outcome variable. We removed disincentives (e.g., for selective tracking) by rewarding the choice of sustainable means of transportation, without penalizing undesired behavior. 

Given all this, our main result is somewhat surprising: Whereas the lack of impact from the communication treatments aligns well with earlier findings on suasoric measures in mobility (see Section \ref{literature}), it is puzzling that even our monetary incentives do not seem to influence mobility behavior at all.

However, our result is not without interpretation---it does not appear to simply stem from methodological shortcomings, such as an insufficient number of participants, but rather suggests that treatment effects of a meaningful size were indeed absent. Instead, it suggests either (or both) of the following two explanations.

The first is to consider the findings at face value. The approach to promoting sustainable mobility behavior may simply not be effective within our sample. After all, divided among the 195~participants, the prize sum of CHF\,3,150 amounts to a mere CHF\,16.15 per participant, or CHF\,0.26 per day. This is a negligible fraction of mobility expenses in Switzerland (including opportunity costs), and commonly accepted price elasticities suggest that no detectable change in behavior should be expected. Particularly within the short two-month time frame, the necessary investments (such as e-bike purchases or public transportation subscriptions) may be unrealistic. Moreover, even when improvements are feasible, intentions often fail to translate into actions: active decision-making is frequently overridden by habits \citep[see][]{ouellette1998habit}, and the formation of new habits typically requires triggers beyond the scope of a mobility competition.

Alternatively, participants may have been persuadable to change their behavior, and this rather serves as a lesson in how \emph{not} to incentivize sustainable mobility. In addition to prolonging the duration of the experiment, there are numerous possibilities that could have increased and maintained engagement. Potential measures include well-established gamification elements such as leaderboards, badges, and the pursuit of personalized goals.  
%\textcolor{red}{@Noah: Eventuell könnte man hier strategisch nochmals das eine oder andere Papier zitieren. Mit der Hoffnung einem Reviewer zu zeigen: Wir machen ein schlechtes Beispiel und du bist der tollste. Ev. strategisch aus "International Journal of Sustainable Transportation" ...} 
Interim communication could have been more closely tailored to participants' contexts\footnote{A laudable example of this approach is the study by \cite{dos2022different}, which divides participants into distinct target groups, demonstrating that these groups differ significantly in their acceptance of mobile applications that use persuasive strategies.}, and monetary incentives could have been made more salient, for instance, by offering more immediate rewards.

Concerning the chosen treatments, it is debatable whether the definition of our outcome variable (see Subsection \ref{scores}) adequately aligns with participants' intuition about the intended goal. In future adaptations of a comparable competition, it might prove more effective to choose an even more straightforward measure of sustainable mobility (such as CO\textsubscript{2} emissions) and treat fairness and cheating safeguards as secondary concerns. A simplification of the rules (such as exclusively focusing on bicycle trips) might also increase memorability and thus engagement.

Likewise, it remains unclear whether our wording regarding the incentives treatments, as exemplified in Excerpts \ref{ex:constant_injunctive} through \ref{ex:no_descriptive} in Appendix \ref{app:excerpts}, was sufficiently comprehensible. Thus, our lesson includes a strong recommendation to pretest communication strategies. However, this should go beyond gauging understandability; it could also help in determining the most compelling arguments to change mobility behavior. Besides social norms and economic reasoning, possible motivations could also include moral arguments, personal norms, or health considerations.

Ultimately, our null result is itself a significant finding, as it offers its own lessons for both researchers and policymakers. We not only suggest rigorously testing interventions before implementing them on a larger scale, but also advocate for redirecting resources to more impactful measures when results are less promising. As beloved as gamified approaches are, when it comes to making meaningful progress towards sustainable mobility, technology, infrastructure, and the frowned-upon push measures may simply be necessary.

	\newpage
	\begin{spacing}{1.0}	
	\bibliographystyle{econometrica}
	\bibliography{MobilityChallenge.bib}
	\end{spacing}
	\newpage
	
\begin{appendix}
		
		\numberwithin{equation}{section}
		\counterwithin{figure}{section}
		\counterwithin{table}{section}
		\noindent \textbf{\LARGE Appendices}
	
\section{Additional Figures}\label{app:figures}
\FloatBarrier

\begin{figure}[H] 
	\begin{tikzpicture}[x=1pt,y=1pt]
	\definecolor{fillColor}{RGB}{255,255,255}
	\path[use as bounding box,fill=fillColor,fill opacity=0.00] (0,0) rectangle (433.62,216.81);
	\begin{scope}
		\path[clip] ( 36.11, 30.69) rectangle (428.12,194.15);
		\definecolor{drawColor}{gray}{0.92}
		
		\path[draw=drawColor,line width= 0.3pt,line join=round] ( 36.11, 51.63) --
		(428.12, 51.63);
		
		\path[draw=drawColor,line width= 0.3pt,line join=round] ( 36.11, 78.65) --
		(428.12, 78.65);
		
		\path[draw=drawColor,line width= 0.3pt,line join=round] ( 36.11,105.66) --
		(428.12,105.66);
		
		\path[draw=drawColor,line width= 0.3pt,line join=round] ( 36.11,132.68) --
		(428.12,132.68);
		
		\path[draw=drawColor,line width= 0.3pt,line join=round] ( 36.11,159.70) --
		(428.12,159.70);
		
		\path[draw=drawColor,line width= 0.3pt,line join=round] ( 36.11,186.72) --
		(428.12,186.72);
		
		\path[draw=drawColor,line width= 0.3pt,line join=round] ( 95.51, 30.69) --
		( 95.51,194.15);
		
		\path[draw=drawColor,line width= 0.3pt,line join=round] (187.57, 30.69) --
		(187.57,194.15);
		
		\path[draw=drawColor,line width= 0.3pt,line join=round] (279.63, 30.69) --
		(279.63,194.15);
		
		\path[draw=drawColor,line width= 0.3pt,line join=round] (368.72, 30.69) --
		(368.72,194.15);
		
		\path[draw=drawColor,line width= 0.6pt,line join=round] ( 36.11, 38.12) --
		(428.12, 38.12);
		
		\path[draw=drawColor,line width= 0.6pt,line join=round] ( 36.11, 65.14) --
		(428.12, 65.14);
		
		\path[draw=drawColor,line width= 0.6pt,line join=round] ( 36.11, 92.15) --
		(428.12, 92.15);
		
		\path[draw=drawColor,line width= 0.6pt,line join=round] ( 36.11,119.17) --
		(428.12,119.17);
		
		\path[draw=drawColor,line width= 0.6pt,line join=round] ( 36.11,146.19) --
		(428.12,146.19);
		
		\path[draw=drawColor,line width= 0.6pt,line join=round] ( 36.11,173.21) --
		(428.12,173.21);
		
		\path[draw=drawColor,line width= 0.6pt,line join=round] ( 53.93, 30.69) --
		( 53.93,194.15);
		
		\path[draw=drawColor,line width= 0.6pt,line join=round] (137.08, 30.69) --
		(137.08,194.15);
		
		\path[draw=drawColor,line width= 0.6pt,line join=round] (238.06, 30.69) --
		(238.06,194.15);
		
		\path[draw=drawColor,line width= 0.6pt,line join=round] (321.21, 30.69) --
		(321.21,194.15);
		
		\path[draw=drawColor,line width= 0.6pt,line join=round] (416.24, 30.69) --
		(416.24,194.15);
		\definecolor{fillColor}{RGB}{190,190,190}
		
		\path[fill=fillColor,fill opacity=0.30] ( 56.90, 30.69) rectangle ( 62.84,194.15);
		
		\path[fill=fillColor,fill opacity=0.30] ( 62.84, 30.69) rectangle ( 68.78,194.15);
		
		\path[fill=fillColor,fill opacity=0.30] ( 98.48, 30.69) rectangle (104.42,194.15);
		
		\path[fill=fillColor,fill opacity=0.30] (104.42, 30.69) rectangle (110.36,194.15);
		
		\path[fill=fillColor,fill opacity=0.30] (140.05, 30.69) rectangle (145.99,194.15);
		
		\path[fill=fillColor,fill opacity=0.30] (145.99, 30.69) rectangle (151.93,194.15);
		
		\path[fill=fillColor,fill opacity=0.30] (181.63, 30.69) rectangle (187.57,194.15);
		
		\path[fill=fillColor,fill opacity=0.30] (187.57, 30.69) rectangle (193.51,194.15);
		
		\path[fill=fillColor,fill opacity=0.30] (223.21, 30.69) rectangle (229.15,194.15);
		
		\path[fill=fillColor,fill opacity=0.30] (229.15, 30.69) rectangle (235.09,194.15);
		
		\path[fill=fillColor,fill opacity=0.30] (264.78, 30.69) rectangle (270.72,194.15);
		
		\path[fill=fillColor,fill opacity=0.30] (270.72, 30.69) rectangle (276.66,194.15);
		
		\path[fill=fillColor,fill opacity=0.30] (306.36, 30.69) rectangle (312.30,194.15);
		
		\path[fill=fillColor,fill opacity=0.30] (312.30, 30.69) rectangle (318.24,194.15);
		
		\path[fill=fillColor,fill opacity=0.30] (347.94, 30.69) rectangle (353.88,194.15);
		
		\path[fill=fillColor,fill opacity=0.30] (353.88, 30.69) rectangle (359.82,194.15);
		
		\path[fill=fillColor,fill opacity=0.30] (389.51, 30.69) rectangle (395.45,194.15);
		
		\path[fill=fillColor,fill opacity=0.30] (395.45, 30.69) rectangle (401.39,194.15);
		\definecolor{drawColor}{RGB}{0,0,0}
		
		\path[draw=drawColor,line width= 0.6pt,line join=round] ( 53.93, 89.45) --
		( 59.87, 92.15) --
		( 65.81, 73.24) --
		( 71.75,154.30) --
		( 77.69,148.90) --
		( 83.63,165.11) --
		( 89.57,178.62) --
		( 95.51,184.02) --
		(101.45,170.51) --
		(107.39,132.68) --
		(113.33,154.30) --
		(119.26,175.91) --
		(125.20,181.32) --
		(131.14,186.72) --
		(137.08,178.62) --
		(143.02,159.70) --
		(148.96,138.09) --
		(154.90,162.41) --
		(160.84,170.51) --
		(166.78,184.02) --
		(172.72,173.21) --
		(178.66,173.21) --
		(184.60,165.11) --
		(190.54,129.98) --
		(196.48,167.81) --
		(202.42,184.02) --
		(208.36,175.91) --
		(214.30,178.62) --
		(220.24,162.41) --
		(226.18,154.30) --
		(232.12,148.90) --
		(238.06, 97.56) --
		(243.99,146.19) --
		(249.93,175.91) --
		(255.87,167.81) --
		(261.81,151.60) --
		(267.75,148.90) --
		(273.69,138.09) --
		(279.63,154.30) --
		(285.57,148.90) --
		(291.51,146.19) --
		(297.45,154.30) --
		(303.39,162.41) --
		(309.33,143.49) --
		(315.27,138.09) --
		(321.21,135.39) --
		(327.15,124.58) --
		(333.09,140.79) --
		(339.03,148.90) --
		(344.97,154.30) --
		(350.91,143.49) --
		(356.85, 97.56) --
		(362.79,132.68) --
		(368.72,132.68) --
		(374.66,157.00) --
		(380.60,157.00) --
		(386.54,154.30) --
		(392.48,148.90) --
		(398.42, 89.45) --
		(404.36,138.09) --
		(410.30,132.68);
		
		\path[draw=drawColor,line width= 0.6pt,dash pattern=on 4pt off 4pt ,line join=round] ( 89.57, 30.69) -- ( 89.57,194.15);
		
		\path[draw=drawColor,line width= 0.6pt,dash pattern=on 4pt off 4pt ,line join=round] (214.30, 30.69) -- (214.30,194.15);
		
		\path[draw=drawColor,line width= 0.6pt,dash pattern=on 4pt off 4pt ,line join=round] (333.09, 30.69) -- (333.09,194.15);
		
		\path[draw=drawColor,line width= 0.6pt,dash pattern=on 4pt off 4pt ,line join=round] (232.12, 30.69) -- (232.12,194.15);
		
		\node[text=drawColor,rotate= 90.00,anchor=base west,inner sep=0pt, outer sep=0pt, scale=  1.10] at ( 85.77, 47.53) {\footnotesize Communication 1};
		
		\node[text=drawColor,rotate= 90.00,anchor=base west,inner sep=0pt, outer sep=0pt, scale=  1.10] at (210.50, 47.53) {\footnotesize Communication 2};
		
		\node[text=drawColor,rotate= 90.00,anchor=base west,inner sep=0pt, outer sep=0pt, scale=  1.10] at (329.29, 47.53) {\footnotesize Communication 3};
		
		\node[text=drawColor,rotate= 90.00,anchor=base west,inner sep=0pt, outer sep=0pt, scale=  1.10] at (228.31, 47.53) {\footnotesize Easter Sunday};
	\end{scope}
	\begin{scope}
		\path[clip] (  0.00,  0.00) rectangle (433.62,216.81);
		\definecolor{drawColor}{gray}{0.30}
		
		\node[text=drawColor,anchor=base east,inner sep=0pt, outer sep=0pt, scale=  0.88] at ( 31.16, 35.09) {140};
		
		\node[text=drawColor,anchor=base east,inner sep=0pt, outer sep=0pt, scale=  0.88] at ( 31.16, 62.11) {150};
		
		\node[text=drawColor,anchor=base east,inner sep=0pt, outer sep=0pt, scale=  0.88] at ( 31.16, 89.12) {160};
		
		\node[text=drawColor,anchor=base east,inner sep=0pt, outer sep=0pt, scale=  0.88] at ( 31.16,116.14) {170};
		
		\node[text=drawColor,anchor=base east,inner sep=0pt, outer sep=0pt, scale=  0.88] at ( 31.16,143.16) {180};
		
		\node[text=drawColor,anchor=base east,inner sep=0pt, outer sep=0pt, scale=  0.88] at ( 31.16,170.18) {190};
	\end{scope}
	\begin{scope}
		\path[clip] (  0.00,  0.00) rectangle (433.62,216.81);
		\definecolor{drawColor}{gray}{0.30}
		
		\node[text=drawColor,anchor=base,inner sep=0pt, outer sep=0pt, scale=  0.88] at ( 53.93, 19.68) {Mar 01};
		
		\node[text=drawColor,anchor=base,inner sep=0pt, outer sep=0pt, scale=  0.88] at (137.08, 19.68) {Mar 15};
		
		\node[text=drawColor,anchor=base,inner sep=0pt, outer sep=0pt, scale=  0.88] at (238.06, 19.68) {Apr 01};
		
		\node[text=drawColor,anchor=base,inner sep=0pt, outer sep=0pt, scale=  0.88] at (321.21, 19.68) {Apr 15};
		
		\node[text=drawColor,anchor=base,inner sep=0pt, outer sep=0pt, scale=  0.88] at (416.24, 19.68) {May 01};
	\end{scope}
\end{tikzpicture}
	\centering \caption{Time series of daily tracked participants (March and April 2024; weekends shaded)} \label{Fig:Timeseries_persons}
\end{figure}
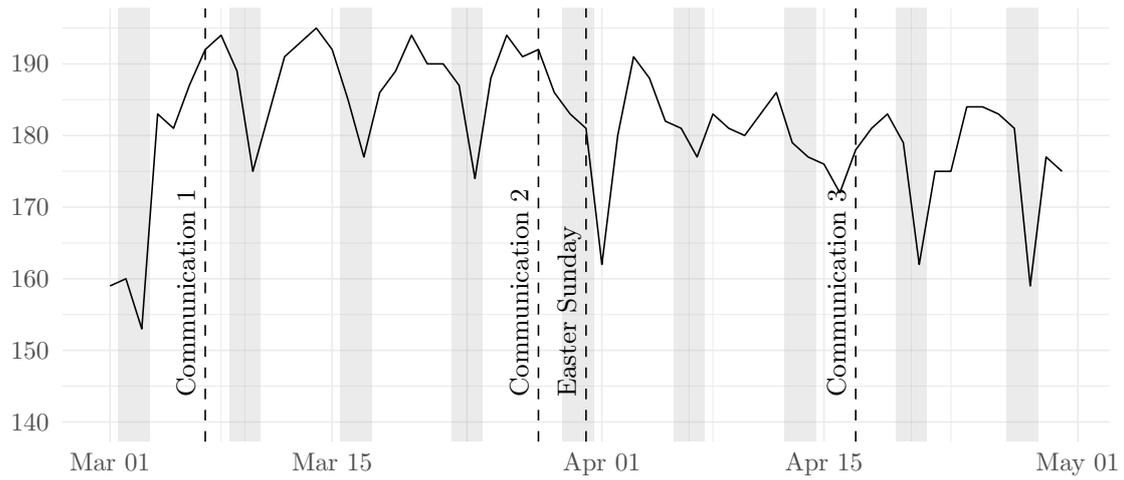

\begin{figure}[H] 
	\input{tikz_daily_points_per_group}
	\centering \caption{Time series of daily points per group (March and April 2024; weekends shaded; incentives treatment: \raisebox{0.5ex}{\rule{14pt}{1pt}} Constant incentives;  \raisebox{0.5ex}{\rule{6pt}{1pt}\hspace{4pt}\rule{5pt}{1pt}} Asymmetric incentives; \raisebox{0.5ex}{\rule{1pt}{1pt}\hspace{2pt}\rule{1pt}{1pt}\hspace{2pt}\rule{1pt}{1pt}\hspace{2pt}\rule{1pt}{1pt}} No incentives)} \label{Fig:Timeseries_point_group}
\end{figure}

\begin{figure}[H] 
	\begin{tikzpicture}[x=1pt,y=1pt]
	\definecolor{fillColor}{RGB}{255,255,255}
	\path[use as bounding box,fill=fillColor,fill opacity=0.00] (0,0) rectangle (433.62,216.81);
	\begin{scope}
		\path[clip] ( 40.51, 30.69) rectangle (428.12,194.15);
		\definecolor{drawColor}{gray}{0.92}
		
		\path[draw=drawColor,line width= 0.3pt,line join=round] ( 40.51, 42.58) --
		(428.12, 42.58);
		
		\path[draw=drawColor,line width= 0.3pt,line join=round] ( 40.51, 71.67) --
		(428.12, 71.67);
		
		\path[draw=drawColor,line width= 0.3pt,line join=round] ( 40.51,100.77) --
		(428.12,100.77);
		
		\path[draw=drawColor,line width= 0.3pt,line join=round] ( 40.51,129.86) --
		(428.12,129.86);
		
		\path[draw=drawColor,line width= 0.3pt,line join=round] ( 40.51,158.96) --
		(428.12,158.96);
		
		\path[draw=drawColor,line width= 0.3pt,line join=round] ( 40.51,188.05) --
		(428.12,188.05);
		
		\path[draw=drawColor,line width= 0.3pt,line join=round] ( 99.24, 30.69) --
		( 99.24,194.15);
		
		\path[draw=drawColor,line width= 0.3pt,line join=round] (190.27, 30.69) --
		(190.27,194.15);
		
		\path[draw=drawColor,line width= 0.3pt,line join=round] (281.30, 30.69) --
		(281.30,194.15);
		
		\path[draw=drawColor,line width= 0.3pt,line join=round] (369.39, 30.69) --
		(369.39,194.15);
		
		\path[draw=drawColor,line width= 0.6pt,line join=round] ( 40.51, 57.13) --
		(428.12, 57.13);
		
		\path[draw=drawColor,line width= 0.6pt,line join=round] ( 40.51, 86.22) --
		(428.12, 86.22);
		
		\path[draw=drawColor,line width= 0.6pt,line join=round] ( 40.51,115.32) --
		(428.12,115.32);
		
		\path[draw=drawColor,line width= 0.6pt,line join=round] ( 40.51,144.41) --
		(428.12,144.41);
		
		\path[draw=drawColor,line width= 0.6pt,line join=round] ( 40.51,173.50) --
		(428.12,173.50);
		
		\path[draw=drawColor,line width= 0.6pt,line join=round] ( 58.13, 30.69) --
		( 58.13,194.15);
		
		\path[draw=drawColor,line width= 0.6pt,line join=round] (140.35, 30.69) --
		(140.35,194.15);
		
		\path[draw=drawColor,line width= 0.6pt,line join=round] (240.19, 30.69) --
		(240.19,194.15);
		
		\path[draw=drawColor,line width= 0.6pt,line join=round] (322.41, 30.69) --
		(322.41,194.15);
		
		\path[draw=drawColor,line width= 0.6pt,line join=round] (416.37, 30.69) --
		(416.37,194.15);
		\definecolor{fillColor}{RGB}{190,190,190}
		
		\path[fill=fillColor,fill opacity=0.30] ( 61.07, 30.69) rectangle ( 66.94,194.15);
		
		\path[fill=fillColor,fill opacity=0.30] ( 66.94, 30.69) rectangle ( 72.81,194.15);
		
		\path[fill=fillColor,fill opacity=0.30] (102.18, 30.69) rectangle (108.05,194.15);
		
		\path[fill=fillColor,fill opacity=0.30] (108.05, 30.69) rectangle (113.92,194.15);
		
		\path[fill=fillColor,fill opacity=0.30] (143.29, 30.69) rectangle (149.16,194.15);
		
		\path[fill=fillColor,fill opacity=0.30] (149.16, 30.69) rectangle (155.03,194.15);
		
		\path[fill=fillColor,fill opacity=0.30] (184.40, 30.69) rectangle (190.27,194.15);
		
		\path[fill=fillColor,fill opacity=0.30] (190.27, 30.69) rectangle (196.14,194.15);
		
		\path[fill=fillColor,fill opacity=0.30] (225.51, 30.69) rectangle (231.38,194.15);
		
		\path[fill=fillColor,fill opacity=0.30] (231.38, 30.69) rectangle (237.25,194.15);
		
		\path[fill=fillColor,fill opacity=0.30] (266.62, 30.69) rectangle (272.49,194.15);
		
		\path[fill=fillColor,fill opacity=0.30] (272.49, 30.69) rectangle (278.36,194.15);
		
		\path[fill=fillColor,fill opacity=0.30] (307.73, 30.69) rectangle (313.60,194.15);
		
		\path[fill=fillColor,fill opacity=0.30] (313.60, 30.69) rectangle (319.47,194.15);
		
		\path[fill=fillColor,fill opacity=0.30] (348.84, 30.69) rectangle (354.71,194.15);
		
		\path[fill=fillColor,fill opacity=0.30] (354.71, 30.69) rectangle (360.58,194.15);
		
		\path[fill=fillColor,fill opacity=0.30] (389.95, 30.69) rectangle (395.82,194.15);
		
		\path[fill=fillColor,fill opacity=0.30] (395.82, 30.69) rectangle (401.69,194.15);
		\definecolor{drawColor}{RGB}{0,0,0}
		
		\path[draw=drawColor,line width= 0.6pt,line join=round] ( 58.13, 39.75) --
		( 64.00, 52.07) --
		( 69.87, 55.04) --
		( 75.75, 40.61) --
		( 81.62, 41.05) --
		( 87.49, 52.76) --
		( 93.37, 48.69) --
		( 99.24, 54.02) --
		(105.11, 56.50) --
		(110.98, 60.99) --
		(116.86, 38.12) --
		(122.73, 41.98) --
		(128.60, 45.47) --
		(134.48, 43.39) --
		(140.35, 59.92) --
		(146.22, 48.59) --
		(152.09, 55.52) --
		(157.97, 44.80) --
		(163.84,118.20) --
		(169.71, 53.79) --
		(175.59, 44.70) --
		(181.46, 53.08) --
		(187.33, 66.39) --
		(193.20, 70.18) --
		(199.08, 43.51) --
		(204.95, 52.55) --
		(210.82, 48.80) --
		(216.70, 55.91) --
		(222.57, 65.16) --
		(228.44, 52.92) --
		(234.32, 61.88) --
		(240.19, 71.41) --
		(246.06, 88.58) --
		(251.93, 46.38) --
		(257.81, 60.25) --
		(263.68,149.76) --
		(269.55, 90.97) --
		(275.43, 72.53) --
		(281.30, 44.14) --
		(287.17, 51.55) --
		(293.04, 49.89) --
		(298.92, 66.69) --
		(304.79, 80.79) --
		(310.66, 95.71) --
		(316.54, 66.71) --
		(322.41, 47.75) --
		(328.28, 55.70) --
		(334.15, 56.87) --
		(340.03, 53.58) --
		(345.90, 44.86) --
		(351.77, 59.90) --
		(357.65, 88.67) --
		(363.52,112.34) --
		(369.39, 67.13) --
		(375.26, 48.56) --
		(381.14, 43.11) --
		(387.01,159.77) --
		(392.88, 63.25) --
		(398.76, 73.07) --
		(404.63,186.72) --
		(410.50, 42.24);
		
		\path[draw=drawColor,line width= 0.6pt,dash pattern=on 4pt off 4pt ,line join=round] ( 93.37, 30.69) -- ( 93.37,194.15);
		
		\path[draw=drawColor,line width= 0.6pt,dash pattern=on 4pt off 4pt ,line join=round] (216.70, 30.69) -- (216.70,194.15);
		
		\path[draw=drawColor,line width= 0.6pt,dash pattern=on 4pt off 4pt ,line join=round] (334.15, 30.69) -- (334.15,194.15);
		
		\path[draw=drawColor,line width= 0.6pt,dash pattern=on 4pt off 4pt ,line join=round] (234.32, 30.69) -- (234.32,194.15);
		
		\node[text=drawColor,rotate= 90.00,anchor=base west,inner sep=0pt, outer sep=0pt, scale=  1.10] at ( 89.56, 95.63) {\footnotesize Communication 1};
		
		\node[text=drawColor,rotate= 90.00,anchor=base west,inner sep=0pt, outer sep=0pt, scale=  1.10] at (212.89, 95.63) {\footnotesize Communication 2};
		
		\node[text=drawColor,rotate= 90.00,anchor=base west,inner sep=0pt, outer sep=0pt, scale=  1.10] at (330.35, 95.63) {\footnotesize Communication 3};
		
		\node[text=drawColor,rotate= 90.00,anchor=base west,inner sep=0pt, outer sep=0pt, scale=  1.10] at (230.51, 95.63) {\footnotesize Easter Sunday};
	\end{scope}
	\begin{scope}
		\path[clip] (  0.00,  0.00) rectangle (433.62,216.81);
		\definecolor{drawColor}{gray}{0.30}
		
		\node[text=drawColor,anchor=base east,inner sep=0pt, outer sep=0pt, scale=  0.88] at ( 35.56, 54.10) {1,000};
		
		\node[text=drawColor,anchor=base east,inner sep=0pt, outer sep=0pt, scale=  0.88] at ( 35.56, 83.19) {2,000};
		
		\node[text=drawColor,anchor=base east,inner sep=0pt, outer sep=0pt, scale=  0.88] at ( 35.56,112.29) {3,000};
		
		\node[text=drawColor,anchor=base east,inner sep=0pt, outer sep=0pt, scale=  0.88] at ( 35.56,141.38) {4,000};
		
		\node[text=drawColor,anchor=base east,inner sep=0pt, outer sep=0pt, scale=  0.88] at ( 35.56,170.47) {5,000};
	\end{scope}
	\begin{scope}
		\path[clip] (  0.00,  0.00) rectangle (433.62,216.81);
		\definecolor{drawColor}{gray}{0.30}
		
		\node[text=drawColor,anchor=base,inner sep=0pt, outer sep=0pt, scale=  0.88] at ( 58.13, 19.68) {Mar 01};
		
		\node[text=drawColor,anchor=base,inner sep=0pt, outer sep=0pt, scale=  0.88] at (140.35, 19.68) {Mar 15};
		
		\node[text=drawColor,anchor=base,inner sep=0pt, outer sep=0pt, scale=  0.88] at (240.19, 19.68) {Apr 01};
		
		\node[text=drawColor,anchor=base,inner sep=0pt, outer sep=0pt, scale=  0.88] at (322.41, 19.68) {Apr 15};
		
		\node[text=drawColor,anchor=base,inner sep=0pt, outer sep=0pt, scale=  0.88] at (416.37, 19.68) {May 01};
	\end{scope}
\end{tikzpicture}
	\centering \caption{Time series of daily CO$_2$ emissions in kg (March and April 2024; weekends shaded)} \label{Fig:Timeseries_CO2}
\end{figure}
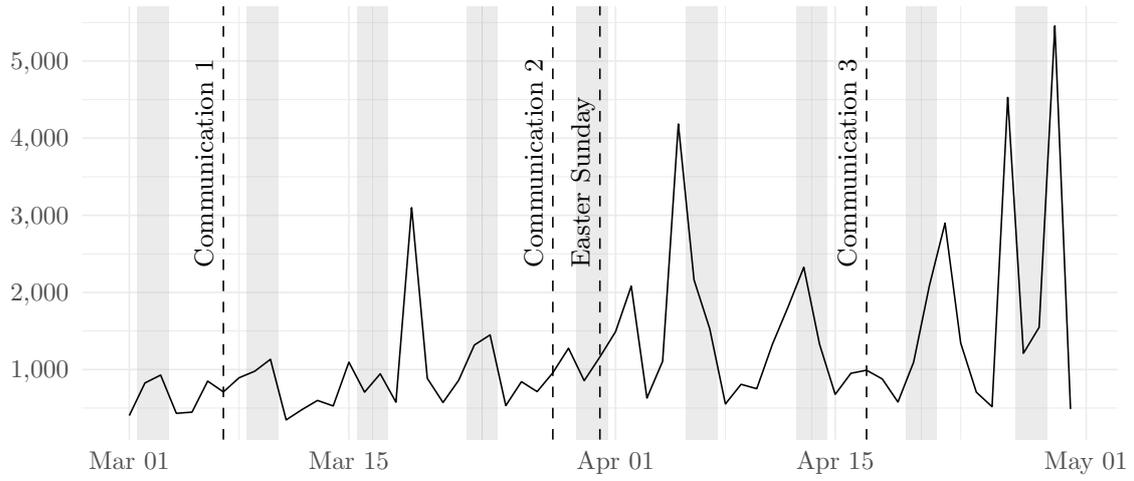

\begin{figure}[H] 
	\input{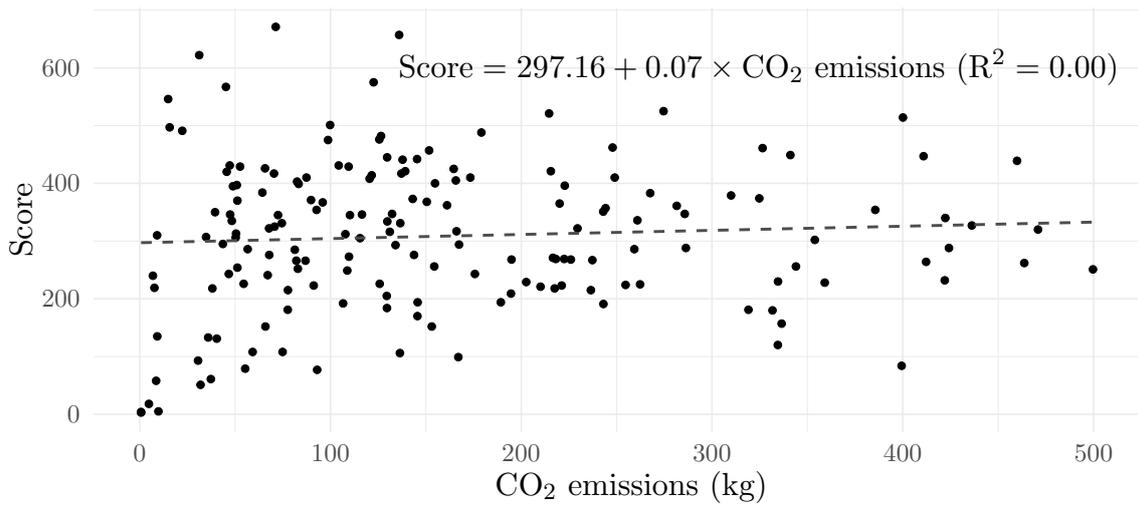}
	\centering \caption{Scatterplot of scores vs. CO$_2$ emissions (March and April 2024)} \label{Fig:Scatter_Points_CO2}
\end{figure}
	
\clearpage
	
\section{Additional Tables}\label{app:tables}
\FloatBarrier
%\begin{table}[H]
%	\centering
%	\footnotesize
%	\singlespacing	
%	\begin{threeparttable}
%	\caption{CO\textsubscript{2} equivalents factors per passenger kilometer} \label{tab:mobitool}
%	\begin{tabular}{ll}
%		\toprule
%		Means of transport                                 & CO\textsubscript{2} equivalents factors  \\ \midrule
%		Passenger   car: diesel, gasoline (fleet average) & 186.4 grams                                                                             \vspace{7pt}\\
%		Passenger   car: battery electric (fleet average) & 89.8 grams                                                                             \vspace{7pt}\\
%		Bicycle                                           & 5.6  grams                                                                             \vspace{7pt}\\
%		Train, regional transportation, s-rail & 8.2  grams                                                                             \vspace{7pt}\\
%		By foot                                           & 0.0 grams \\ \bottomrule                                                                             
%	\end{tabular}
%	\begin{tablenotes}
%		\item Source: \cite{mobitool2023}.
%	\end{tablenotes}
%\end{threeparttable}
%\end{table}

\begin{table}[H]
	\centering
	\footnotesize
	\singlespacing	
	\begin{threeparttable}
		\caption{CO\textsubscript{2} equivalents factors per passenger kilometer} \label{tab:mobitool}
		\begin{tabular}{ll}
			\toprule
			Means of transport                                 & CO\textsubscript{2} equivalents factors  \\ \midrule
			Airplane & 263.0 grams                                                                             \vspace{7pt}\\
			Car, other & 186.4 grams                                                                             \vspace{7pt}\\
			Motorbike & 163.6 grams                                                                             \vspace{7pt}\\
			Ferry & 161.3 grams                                                                             \vspace{7pt}\\
			Bus & 133.8 grams                                                                             \vspace{7pt}\\
			Electric car & 89.8 grams                                                                             \vspace{7pt}\\
			Coach & 46.5 grams                                                                             \vspace{7pt}\\
			E-scooter & 45.7 grams                                                                             \vspace{7pt}\\
			Cable car, subway, tramway & 42.8 grams                                                                             \vspace{7pt}\\
			E-bike                                           & 11.3  grams                                                                             \vspace{7pt}\\
			Rapid transit railway, regional train, train & 7.0  grams                                                                             \vspace{7pt}\\
			Bicycle, bikesharing                                           & 5.6  grams                                                                             \vspace{7pt}\\
			Walking                                           & 0.0 grams \\ \bottomrule                                                                             
		\end{tabular}
		\begin{tablenotes}
			\item Source: \cite{mobitool2023}.
			\item Bikesharing, e-bike, and electric car are only applied if manually overwritten by participant (not relevant for scores).
		\end{tablenotes}
	\end{threeparttable}
\end{table}
	
\begin{table}[H]
	\centering
	\footnotesize
	\singlespacing	
	\begin{threeparttable}
		\caption{Regression of scores on incentives (binary) and communication (binary)}\label{tab:effects_simple}
		\begin{tabular}{p{3cm}cccc}           
			\toprule
			& (1) & (2) & (3) & (4) \\ \midrule 
			\multirow{2}{3cm}{(Intercept)}                           & \num{312.58}***                & \num{306.88}***                & \num{311.24}***                & \num{328.43}***                \\ 
			& [\num{271.69}, \num{353.46}] & [\num{268.20}, \num{345.55}] & [\num{263.71}, \num{358.76}] & [\num{266.15}, \num{390.71}] \vspace{7pt}\\ 
			\multirow{2}{3cm}{Student}                          & \num{-10.26}                   & \num{-10.22}                   & \num{-10.44}                   & \num{-12.00}                   \\
			& [\num{-48.58}, \num{28.05}]  & [\num{-48.65}, \num{28.21}]  & [\num{-48.98}, \num{28.11}]  & [\num{-50.75}, \num{26.74}]  \vspace{7pt}\\ 
			\multirow{2}{3cm}{Incentives}                       & \num{-6.22}                    &                                  & \num{-6.16}                    & \num{-29.84}                   \\
			& [\num{-44.95}, \num{32.52}]  &                                  & [\num{-45.01}, \num{32.69}]  & [\num{-97.50}, \num{37.81}]  \vspace{7pt}\\ 
			\multirow{2}{3cm}{Communication}                    &                                  & \num{2.33}                     & \num{2.16}                     & \num{-21.68}                   \\
			&                                  & [\num{-36.23}, \num{40.89}]  & [\num{-36.51}, \num{40.83}]  & [\num{-89.52}, \num{46.16}] \vspace{7pt} \\
			\multirow{2}{3cm}{Incentives × Communication} &                                  &                                  &                                  & \num{35.43}                    \\
			&                                  &                                  &                                  & [\num{-47.39}, \num{118.24}] \vspace{7pt}\\
			R$^2$                                    & \num{0.002}                     & \num{0.001}                     & \num{0.002}                     & \num{0.006}                     \vspace{7pt}\\
			Adjusted R$^2$                                      & \num{-0.008}                     & \num{-0.009}                     & \num{-0.014}                     & \num{-0.015}                     \\
			\bottomrule
		\end{tabular}
		\begin{tablenotes}[flushleft]
			\setlength{\itemsep}{2pt} % Adjust item separation
			\item Number of observations: 195.
			\item Incentives = 1: Incentives $\in$ \{Constant incentives, Asymmetric incentives\}.
			\item Communication = 1: Communication $\in$ \{Injunctive norm, Descriptive norm\}.
			\item * p < 0.1, ** p < 0.05, *** p < 0.01.
		\end{tablenotes}
	\end{threeparttable} 
\end{table}	
\clearpage

\section{Communication Excepts (Translated from German)}
\sethlcolor{lightgray}\singlespacing	
\addcontentsline{toc}{subsection}{Communication Excerpts}
\emph{Notes: Shading, added in Excerpts \ref{ex:constant_injunctive} to \ref{ex:no_descriptive} to highlight key passages, was not present in the original communication. The translation was conducted with the assistance of ChatGPT.}
\label{app:excerpts}
\FloatBarrier
\begin{enumerate}[leftmargin=*] 	
	\item \textbf{Invitation e-mail (treatment: \emph{Asymmetric incentives})}
	\label{ex:start}

	Dear Participant of the HSLU Mobility Challenge 2024,
	
	In just a few days, it will begin. From March 1\textsuperscript{st} to April 30\textsuperscript{th}, 2024, you can collect points by using environmentally friendly modes of transportation as often as possible. This will provide you with feedback on your mobility behavior and give you the chance to win attractive prizes.
	
	As a first step, download the MotionTag app from the App Store or Play Store. Then, open the app and register using your HSLU e-mail address. Choose your own password and select the project "HSLUMC24". You will then receive an e-mail with an activation link, along with the participation terms and data protection information. After confirming the activation link, you can log into the MotionTag app. During the "first steps," allow MotionTag to access your device’s location even when the app is running in the background. If you encounter any issues during registration in the app, please contact \href{mailto:app-support@motion-tag.com}{app-support@motion-tag.com}.
	
	By recording trips for the first time by March 1\textsuperscript{st}, you will already qualify for the first prize draw, where you can win an upcycled bicycle worth CHF\,800 and four public transportation vouchers worth CHF\,100 each.
	
	Starting March 1\textsuperscript{st}, you will earn two points for each "trip" with the main mode of transportation being "slow traffic" (on foot or by bike/e-bike) and one point for each trip with the main mode of transportation being "public transportation."
	
	The challenge is held in three groups. Depending on the group, the points you collect will influence the prizes to be won at the end of the challenge. You have already been randomly assigned to a group.
	
	Among the 88 participants in your group, the three with the most points at the end of the challenge will receive the following prizes:
	
	\begin{itemize}
		\item 1\textsuperscript{st} place: Mobility voucher (public transportation, car-sharing, etc.) worth CHF\,600
		\item 2\textsuperscript{nd} place: Mobility voucher worth CHF\,300
		\item 3\textsuperscript{rd} place: Mobility voucher worth CHF\,150
	\end{itemize}
	Throughout the challenge, you will be informed of your point total three times. Further details on collecting points and the prizes can also be found in the participation terms (§ 5 Rewards).
	
	Please note that MOTIONTAG GmbH, based in Potsdam (Germany), may continue to collect and process your data according to MOTIONTAG’s data protection guidelines after April 30th, 2024, if you continue to use the app. To prevent this, you must deactivate the app (e.g., by uninstalling it). If you wish to have your personal data deleted, you can request this by sending an e-mail to \href{mailto:app-support@motion-tag.com}{app-support@motion-tag.com} (see data protection guidelines).
	
	We wish you much success and enjoyment in the HSLU Mobility Challenge 2024.
	
	Silvio, Hannes, Chiara, and Noah
	
	\pagebreak[2]
	
	\item \textbf{Intermin communication with treatments \emph{Constant incentives} and \emph{Injunctive norm}}
	\label{ex:constant_injunctive}
	
	Hello \censor{dumbass},
	
	The HSLU Mobility Challenge 2024 has been underway for almost a month now.
	
	We would like to inform you about your current point standing.
	
	You have accumulated 92 points. On average, participants in your group have accumulated 132 points. \hl{You can do better!} The competition runs until April 30\textsuperscript{th}.
	
	As a reminder, the following prizes will be raffled among the participants in your group at the end of the challenge:
	
	\begin{itemize}
	\item 1\textsuperscript{st} place: Mobility voucher (public transportation, car-sharing, etc.) worth CHF\,600
	\item 2\textsuperscript{nd} place: Mobility voucher worth CHF\,300
	\item 3\textsuperscript{rd} place: Mobility voucher worth CHF\,150
	\end{itemize}
	
	\hl{In your group, the prizes will be raffled based on the points accumulated: For each of your points, you will receive one ticket at the end of the competition. All tickets have the same chance of winning. Consequently, the more points you collect during the competition, the higher your chances of winning.}
	
	We wish you continued success and enjoyment in the HSLU Mobility Challenge.
	
	Best regards,
	
	Silvio, Hannes, Chiara, and Noah

	\item \textbf{Intermin communication with treatments \emph{Asymmetric incentives} and \emph{No norm}}
	\label{ex:asymmetric_no}

	Hello \censor{Silvio},
	
	The HSLU Mobility Challenge 2024 has been underway for almost a month now.
	
	We would like to inform you about your current point standing.
	
	You have accumulated 92 points. The competition runs until April 30\textsuperscript{th}.
	
	\hl{As a reminder: The three participants in your group with the most points at the end of the challenge will receive the following prizes:}
	
	\begin{itemize}
		\item 1\textsuperscript{st} place: Mobility voucher (public transportation, car-sharing, etc.) worth CHF\,600
		\item 2\textsuperscript{nd} place: Mobility voucher worth CHF\,300
		\item 3\textsuperscript{rd} place: Mobility voucher worth CHF\,150
	\end{itemize}
	
	We wish you continued success and enjoyment in the HSLU Mobility Challenge.
	
	Best regards,
	
	Silvio, Hannes, Chiara, and Noah		

	\pagebreak[4]

	\item \textbf{Intermin communication with treatments \emph{No incentives} and \emph{Descriptive norm}}
	\label{ex:no_descriptive}

	Hello \censor{Hannes},
	
	The HSLU Mobility Challenge 2024 has been underway for almost a month now.
	
	We would like to inform you about your current point standing.
	
	You have accumulated 92 points. \hl{On average, participants in your group have accumulated 132 points.} The competition runs until April 30\textsuperscript{th}.
	
	As a reminder, the following prizes will be raffled among the participants in your group at the end of the challenge:
	
	\begin{itemize}
		\item 1\textsuperscript{st} place: Mobility voucher (public transportation, car-sharing, etc.) worth CHF\,600
		\item 2\textsuperscript{nd} place: Mobility voucher worth CHF\,300
		\item 3\textsuperscript{rd} place: Mobility voucher worth CHF\,150
	\end{itemize}
	
	\hl{However, in your group, the points have no impact on the chances of winning.}
	
	We wish you continued success and enjoyment in the HSLU Mobility Challenge.
	
	Best regards,
	
	Silvio, Hannes, Chiara, and Noah
	
	\item \textbf{Figures from mobility reports}\label{ex:mobility_report}
	\begin{figure}[H]
		\includegraphics[scale=.6]{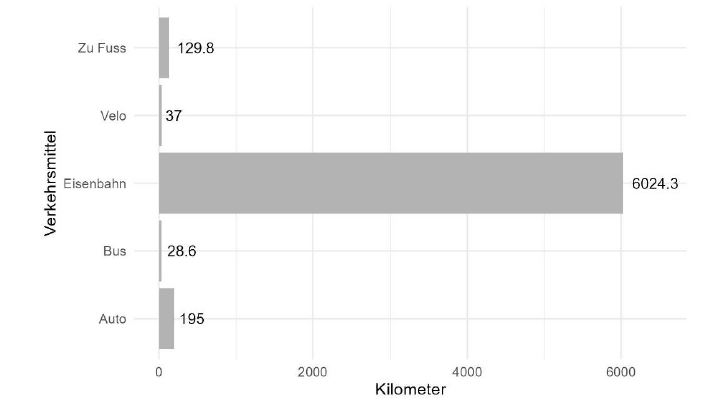}
		\centering \caption{Kilometers traveled by main means of transportation (sources: MotionTag)} \label{fig:kilometer}
	\end{figure}
	\begin{figure}[H]
		\includegraphics[scale=.5]{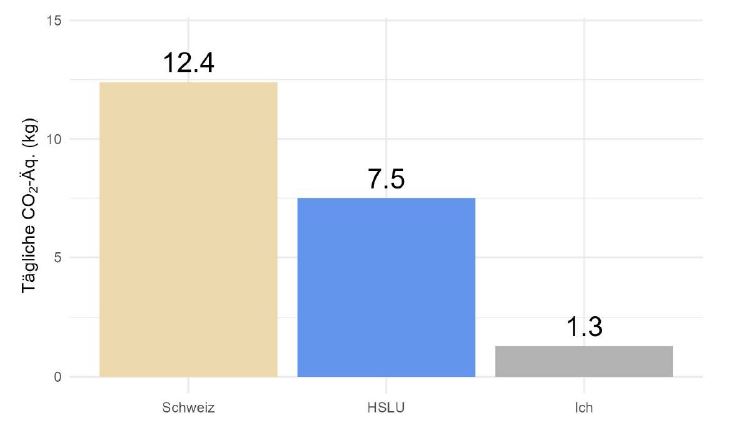}
\centering \caption{Average daily CO\textsubscript{2} emissions (sources: MotionTag, \cite{bfs_are2023}); note that "Switzerland" refers to all age categories and seasons, whereas "Personal" and "HSLU" refer to the conditions and settings of the competition} \label{fig:CO2}
	\end{figure}
	
\end{enumerate}
	\end{appendix}
\end{document}